\documentclass[a4paper,fleqn]{cas-dc}

\usepackage[numbers]{natbib}

\usepackage{graphicx,subfig,epsfig}
\usepackage{amssymb,amsmath,amsbsy,bm,bbm}
\usepackage{algorithm,algpseudocode,theorem}
\usepackage{natbib}
\usepackage{multirow}

\begin{document}
\let\WriteBookmarks\relax
\def\floatpagepagefraction{1}
\def\textpagefraction{.001}

\shorttitle{Modeling of CAR T-cells and Glioma interaction}    

\shortauthors{R. Li et al.}  

\title[mode = title]{Modeling interaction of Glioma cells and CAR T-cells  considering multiple CAR T-cells bindings}  



%

\author[1]{Runpeng Li}
\credit{Analysis, Investigation, Visualization, Software, Writing--original draft preparation}

\author[2]{Prativa Sahoo}
\credit{Data curation, Interpretation, Writing--review and editing}

\author[3]{Dongrui Wang}
\credit{Data curation}

\author[1,4]{Qixuan Wang}
\credit{Methodology, Writing--review and editing}

\author[3]{Christine E. Brown}
\credit{Data curation, Writing--review and editing}

\author[2]{Russell C. Rockne}
\credit{Data curation, Interpretation, Writing--review and editing}

\author[1,4]{Heyrim Cho}
\credit{Conceptualization, Methodology, Analysis, Visualization, Writing--review and editing}
\ead{heyrimc@ucr.edu}
\cormark[1]
\cortext[cor1]{Corresponding author}





\affiliation[1]{organization={Department of Mathematics, University of California Riverside},
            addressline={900 University Ave.}, 
            city={Riverside},
            postcode={92521}, 
            state={CA},
            country={USA}}

\affiliation[2]{organization={Division of Mathematical Oncology, Department of Computational and Quantitative Medicine, Beckman Research Institute, City of Hope National Medical Center},
            addressline={1500 E Duarte Rd.}, 
            city={Duarte},
            postcode={91010}, 
            state={CA},
            country={USA}}
            
\affiliation[3]{organization={Department of Hematology \& Hematopoietic Cell Transplantation, Beckman Research Institute, City of Hope National Medical Center},
            addressline={1500 E Duarte Rd.}, 
            city={Duarte},
            postcode={91010}, 
            state={CA},
            country={USA}}

\affiliation[4]{organization={Interdisciplinary Center for Quantitative Modeling in Biology, University of California Riverside},
            addressline={900 University Ave.}, 
            city={Riverside},
            postcode={92521}, 
            state={CA},
            country={USA}}
            
            




\cortext[1]{Corresponding author}



\begin{abstract}
Chimeric antigen receptor (CAR) T-cell based immunotherapy has shown its potential in treating blood cancers, and its application to solid tumors is currently being extensively investigated. For glioma brain tumors, various CAR T-cell targets include IL13R$\alpha$2, EGFRvIII, HER2, EphA2, GD2, B7-H3, and chlorotoxin. In this work, we are interested in developing a mathematical model of IL13R$\alpha$2 targeting CAR T-cells for treating glioma. We focus on extending the work of Kuznetsov et al. (1994) by considering binding of multiple 
CAR T-cells  to a single glioma cell, and  the dynamics of these multi-cellular conjugates. Our  model more accurately describes experimentally observed CAR T-cell killing assay data than a model which does not consider cell binding. Moreover, we derive conditions in the CAR T-cell expansion rate  that determines  treatment success or failure. Finally, we show that our model captures distinct CAR T-cell killing dynamics at low, medium, and high antigen receptor densities in patient-derived brain tumor cells. 
\end{abstract}



\maketitle 
\section{Introduction}
\label{intro}
Adoptive cell-based immunotherapy has shown to be successful in treating patients with cancer. In particular, Chimeric Antigen Receptor T cell (CAR T-cell) therapy is one of the adoptive immunotherapies that has been successful in clinical and pre-clinical models, and has been FDA-approved since 2017 \cite{Rotolo2016,Mohanty2019}. In this therapy, a patient or donor's T cells is collected, and genetically engineered to express a receptor specific to an antigen found on cancer cells, thus improving the ability of T-cells to eradicate the target cancer cells. Finally, these CAR T-cells are cultured to large numbers, then introduced back to the patient \cite{Paucek2019,Hinrichs2016}. CAR T-cell therapy has shown its potential in blood cancer, and to  solid tumors \cite{June2018,Paucek2019}. However, the success of CAR T-cell therapy for solid tumors has been  challenging due to difficulties in 1) trafficking CAR T-cells into  solid tumors, 2) hostile tumor microenvironment that suppresses T cell activity, and 3) tumor antigen heterogeneity \cite{June2018,marofi2021}.  Moreover, since CAR T-cells mostly exist in bloodstream and lymphatic system, which although makes CAR T-cell therapy a great weapon for hematological tumors (e.g. blood tumor cells), it may be hard for CAR T-cells to penetrate tumor tissue through the vascular endothelium. At last, different cells can infiltrate solid tumors and help to support its growth, which restricts CAR T-cell therapy in the end \cite{marofi2021}.

One of the first mathematical models describing the interaction between immune cells and cancer cells is from Kuznetsov et al. (1994) \cite{Kuznetsov1994}, which is a dynamical system model with two populations, tumor cells and cytotoxic T cells, or T lymphocytes. The model can describe the formation of a tumor dormant state and evasion of the immune system. A subsequent model developed by Kirschner and Panetta (1998) \cite{Kirschner1998}  considered the cytokine interleukin-2 in addition to the dynamics between tumor cells and immune effector cells, and was able to model short-term tumor oscillations as well as long-term tumor relapses. In order to model persistent oscillations that were observed in immune systems, periodic treatment and time delay was added to the model in \cite{Costa2003}, and stability analysis of the model was done in \cite{DOnofrio2008}. Thereafter, the later built models were improved by adding new types of cells, such as natural killer cells, normal cells, and different kinds of cytokines \cite{DePillis2003,Moore2004}. These models not only capture tumor immune escape, but also explain multiple equilibrium phases of coexisting immune cells and cancer cells. Other than dynamical system models, spatial models that describe the spatial temporal interaction are developed as well. \cite{Chaplain2006} develops a spatial temporal version of \cite{Kuznetsov1994} using partial differential equations, \cite{Mallet2006} develops a hybrid cellular automata-partial differential equation model, and \cite{Macklin2018} develops a hybrid off-lattice agent-based and partial differential equation model.

The surge of clinical trials and the success of CAR T-cell therapy also drew a lot of interest in mathematical modeling of CAR T-cell therapy. This includes modeling CD19 CAR T-cell therapy targeting acute lymphoblastic leukemia in \cite{Mostolizadeh2018} as a dynamical system, which includes healthy B cell populations and circulating lymphocytes. Because of the lack of data calibration, later study in \cite{Hardiansyah2019} further showed relationships between CAR T-cell doses and diseases burden by the observed clinical data. In order to study cytokine release syndrome, which is one of the primary side effects of CAR T-cell therapy, another dynamical system of nine cytokines responding to CAR T-cell therapy was proposed in \cite{Hopkins2018} as well. More recent work in \cite{kimmel2021CART} attempted to understand the dynamics of CAR T-cell therapy by considering not only tumor cells and CAR T-cells but also normal T cells. Meanwhile, the model introduced in \cite{barros2020CART} includes long-term memory CAR T-cells, which are produced by memory pool formation of effector CAR T-cells. The corresponding stability analysis of this model was done in \cite{barros2021CART}.

\begin{figure}[ht!]
  \includegraphics[width=8.5cm]{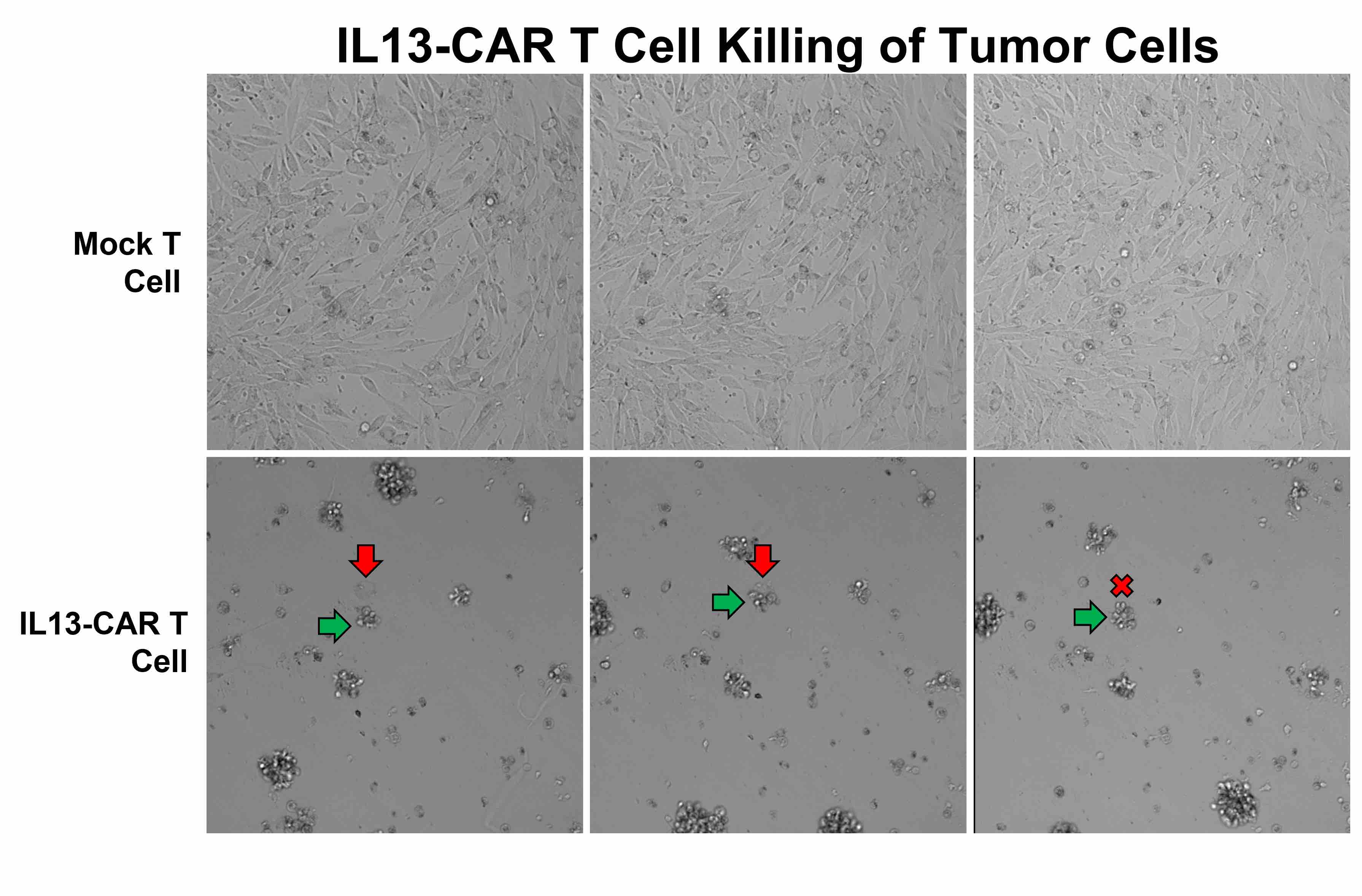}
\caption{IL13-CAR T-cell killing of brain tumor cells. The first row corresponds to when glioma cells are mixed with mock-T cells, and the second row is when mixed with IL13-CAR T-cells. It shows the process of  glioma cell (red arrow) being eliminated by the CAR T-cells (green arrow).}
\label{fig:experiment1}
\end{figure}

Here we develop a mathematical model of glioma cells and CAR T-cells interaction inspired by the experimental data provided in \cite{Sahoo2019}. It studies the interaction between glioma cell lines,  derived from glioblastoma patients undergoing tumor resections at City of Hope \cite{brown2012, brown2016}, and IL13R$\alpha$2 targeting CAR T-cells. As shown in Figure \ref{fig:experiment1}, cells were co-cultured in vitro and images were taken under a light microscope over a 72 hour period. In subsequent experiments, the glioma cells and CAR T-cells are mixed at different ratios (CAR T-cell to  glioma cell ratios of 1:5, 1:10, and 1:20). In addition, glioma cells with different antigen receptor density levels (low, medium, and high) were tested. Throughout the course of experiment, real-time monitoring of glioma population size was performed by using xCELLigence cell analyzer system \cite{moniri2015}, where the size is tracked every 15 minutes. This system quantifies the glioma cell population with a dimensionless number referred to as cell-index (CI). We will refer to this unitless data as glioma tumor size and calibrate the model to this data. 

The rest of the paper is structured as follows. In section \ref{sec:model}, we describe the proposed model, where we assume that one glioma cell may interact with multiple CAR T-cells, and further come up with a system of ODE that describes not only the size of tumor cells and CAR T-cells, but also the size of multiple CAR T-cell binding conjugates. We denote the model with conjugates up to $n$ CAR T-cells binding to glioma cell as the $n$-binding model. The stability analysis of one-binding slow reaction model is presented in section \ref{sec:analysis}, and we provide parameter conditions that guarantee either CAR T-cell treatment success or failure. In section \ref{sec:simulation}, we compare accuracy between the slow reaction and fast reaction one-binding model, and show that the slow reaction model more accurately describes the experimental data. We also simulate the multiple binding slow reaction models from one- to five-binding conjugates, and study the hypotheses in reaction rates that captures the experimental result regarding low to high antigen receptor density levels of glioma cells. Summary of our findings and future work is discussed in section \ref{sec:conclution}.

\section{Mathematical Model}
\label{sec:model}

This section summarizes the mathematical models that we study in this work. We start by deriving the dynamical system model proposed in Kuznetsov et al. (1994) \cite{Kuznetsov1994}, and point out the slow reaction version of the model. Then we extend the model to considering the conjugate of multiple CAR T-cells and glioma cell, as it is motivated from the experiments. We will denote the conjugate that consist of $n$ CAR T-cells and one glioma cell as conjugate $I_n$, and the model that includes the conjugates $I_1$, ..., $I_n$ as the n CAR T-cell binding models, in short, n-binding model.

\subsection{One CAR T-cell bound to one glioma cell  model}\label{sec:model1}

\begin{figure}[ht!]
\centerline{  \includegraphics[width=8cm]{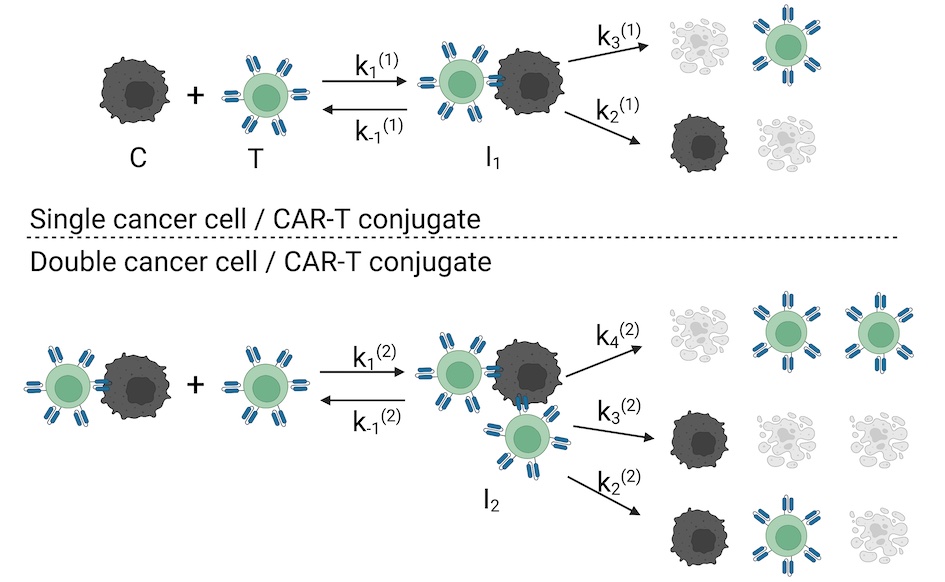}  }   
\caption{Kinetic interaction between glioma cancer cells ($C$) and CAR T-cells ($T$), assuming conjugates $I_1$ of one CAR T-cell to one glioma cell (top) and conjugates $I_2$ of two CAR T-cells to one glioma cell (bottom). } 
\label{fig:interaction}
\end{figure}

Among the mathematical models considering interaction of glioma cells and immune cells, one of the most recognized models is the model from Kuznetsov et al. (1994) \cite{Kuznetsov1994}, modeling the one to one binding of cancer and cytotoxic immune cells. The glioma cells are subject to be attacked by cytotoxic effector cells, e.g. T cells. Here, we consider CAR T-cells instead of the non-adoptive T cells. 
The interaction between the glioma cancer cells $C(t)$ and CAR T-cells $T(t)$ in vitro can be described by the following kinetic scheme in Figure  \ref{fig:interaction} (top), where $I_1(t)$ is conjugate of one CAR T-cell and one glioma cell, $T^\times$ is the inactivated effector cell, and $C^\times$ is the lethally hit glioma cell that are programmed to die. The kinetic parameter $k_1^{(1)}$ describes the rate that the glioma cell binds to the CAR T-cell, and $k_{-1}^{(1)}$ is the rate that the conjugate detaches without damaging the cells. $k_3^{(1)}$ is the rate that the CAR T-cell and glioma cell interaction kills the glioma cell, and  $k_2^{(1)}$ is the rate that damages the CAR T-cell.

In addition to the interaction, the growth dynamics of the glioma cells and CAR T-cells are included in the model, which yields in the following system of differential equation. 
\begin{eqnarray}
\nonumber 
\dot{T} &=& pT\frac{C}{g+C}  - \theta T - k_1^{(1)} C T + k_{-1}^{(1)} I_1  + k_3^{(1)} I_1 \\ 
\dot{C} &=& \rho C ( 1-  \frac{ C}{K}) - k_1^{(1)} C T  + k_{-1}^{(1)} I_1 + k_2^{(1)} I_1 \\  \nonumber 
\dot{I_1} &=& k_1^{(1)} C T - k_{-1} ^{(1)}I_1 - k_3^{(1)} I_1 - k_2^{(1)} I_1.    
\end{eqnarray} 
Here, parameter $\rho$ is the maximal growth rate of the glioma cells assuming logistic growth; parameter $K$ is the maximal carrying capacity of the biological environment for the glioma cells; $p$ is the rate that the CAR T-cells accumulate in the region due to the presence of the tumor, in other words, CAR T-cell expansion rate; $g$ is a concentration of glioma cells $C$ that halves the maximum rate $p$; $\theta$ represents the death rate of the CAR T-cells.
Note that both inactivated cells, $C^\times $ and $T^\times $,  will decay to zero eventually, therefore, we will omit the two equations. We reparameterize the equation as 
\begin{eqnarray}
\label{eq:1bindslow}
\nonumber 
\dot{T} &=& p T \frac{C}{g + C}  - \theta T - k C T + \alpha I_1 \\ 
\dot{C} &=& \rho C \left( 1-  \frac{ C}{K}\right) - k C T  + \beta I_1 \\  \nonumber 
\dot{I_1} &=& k C T -\gamma I_1,    
\end{eqnarray} 
where
\begin{align*}
k &= k_1^{(1)}, \quad \quad \alpha = k_{-1}^{(1)}+k_3^{(1)}\\ \beta &= k_{-1}^{(1)}+k_2^{(1)}, \quad \gamma = k_{-1}^{(1)}+k_2^{(1)}+k_3^{(1)}. 
\end{align*}
We will refer to Eq. \eqref{eq:1bindslow} as the {\em slow reaction} model, since following the dynamics of the conjugate $I_1$ allows the interaction to be in a comparable time scale as the other dynamics, in contrast to the assumption that will be making in the following model.

\subsubsection{One-to-one binding model with fast reaction } 

The model proposed in \cite{Kuznetsov1994} further reduces Eq. \eqref{eq:1bindslow} by a separation of time scale between the dynamics of glioma cells and CAR T-cells compared to the dynamics of the conjugates $I_1$. Assuming that the dynamics of conjugate $I_1$ reaches equilibrium quickly, we can take $\dot{I}_1 = 0$, and we have $k_1^{(1)} C T - k_{-1}^{(1)} I_1 - k_3^{(1)} I_1 - k_2 ^{(1)}I_1=0$, that is, 
\begin{equation*}
I_1 = \frac{k_1^{(1)} }{k_{-1}^{(1)} + k_3^{(1)} + k_2^{(1)}} C T. 
\end{equation*} 
Then, Eq. \eqref{eq:1bindslow} reduces to the Kuznetsov et al. (1994)  model as 
\begin{eqnarray}
&\,&\dot{T}=pT\frac{C}{g+C}-\theta T-mCT \nonumber \\
&\,&\dot{C}=\rho C\left(1-\frac{C}{K}\right)-nCT, 
\label{eq:Kuz}
\end{eqnarray}
where we define  
\begin{equation}
n=\frac{k_1^{(1)}k_3^{(1)}}{k_{-1}^{(1)}+k_2^{(1)}+k_3^{(1)}},\quad m=\frac{k_1^{(1)}k_2^{(1)}}{k_{-1}^{(1)}+k_2^{(1)}+k_3^{(1)}}. 
\label{eq:param_relation}
\end{equation}
We will refer to this model as the {\em fast reaction} model of one CAR T-cell binding case, as opposed to the slow reaction model in Eq. \eqref{eq:1bindslow}. 
The fast reaction scenario can happen, for example, when the glioma cell is mixed with a large number of CAR T-cells or the receptor density levels of glioma cells are high, which may lead to a fast elimination of glioma cells.

\subsection{Two CAR T-cells bound to one glioma cell  model}
The following system, building up on the one-binding models in Eqs. \eqref{eq:1bindslow} and \eqref{eq:Kuz}, will govern the population dynamics for the scenario where up to two CAR T-cells can interact and bind to one glioma cell. As in the diagram of Figure \ref{fig:interaction}, conjugate $I_1$ of one CAR T-cell and one glioma cell can interact with another CAR T-cell ($T$) and compose conjugate $I_2$ of two CAR T-cells and one glioma cell. The kinetic interaction rates are defined similarly as before, where $k_1^{(2)}$ and $k_{-1}^{(2)}$ will describe the binding and detachment rate of cells without damage, and $k_i^{(2)}$ will be the rates of the interaction events. The system of equations can be written as follows. 
\begin{eqnarray}
\dot{T} &=& p T \frac{C}{g + C}  - \theta T - k C T + \alpha I_1  \nonumber \\  
&\,& -k_1^{(2)}I_1 T + (k_{-1}^{(2)} + 2 k_4^{(2)} + k_3^{(2)}) I_2  \\  
\dot{C} &=& \rho C \left( 1-  \frac{ C}{K}\right) - k C T  + \beta I_1 + (k_3^{(2)} + k_2^{(2)}) I_2 \nonumber \\ 
\dot{I_1} &=& k C T - \gamma I_1 -k_1^{(2)} I_1T + k_{-1}^{(2)}I_2   \nonumber \\   
\dot{I_2} &=& k_1^{(2)}I_1 T - (k_{-1}^{(2)}+k_4^{(2)}+k_3^{(2)}+k_2^{(2)})I_2 
 \nonumber 
\label{eq:2bindslow}
\end{eqnarray}
This model follows the dynamics of $I_1$ and $I_2$ conjugates, so we will refer to this as a two-to-one binding slow reaction model. This model describes the scenario that if one CAR T-cell cannot kill a glioma cell immediately, then another CAR T-cell may come to react with the conjugate formed by previous reaction.

\subsubsection{Two-to-one binding model with fast reaction } 

%
Using a similar approach when we derived the Kuznetsov et al. (1994)  model, the two-binding slow reaction system can be reduced to the following two-binding fast reaction system. 
\begin{align} 
\label{eq:2bindfast} 
\dot{T}&=pT\frac{C}{g+C}-\theta T-\frac{k_1^{(1)}k_2^{(1)}}{k_1^{(2)}T+k_2^{(1)}+k_3^{(1)}}CT \nonumber\\
&-\frac{k_1^{(1)}k_1^{(2)}(k_2^{(2)}+2k_3^{(2)})}{(k_2^{(2)}+k_3^{(2)}+k_4^{(2)})(k_1^{(2)}T+k_2^{(1)}+k_3^{(1)})}CT^2 \nonumber\\ 
\dot{C}&=\rho C(1-\frac{C}{K})-\frac{k_1^{(1)}k_3^{(1)}}{k_2^{(1)}+k_3^{(1)}+k_1^{(2)}T}CT \\ 
&-\frac{k_{4}^{(2)}k_1^{(2)}k_1^{(1)}}{(k_2^{(2)}+k_3^{(2)}+k_4^{(2)})(k_2^{(1)}+k_3^{(1)}+k_1^{(2)}T)}CT^2. \nonumber
\end{align} 
Note that we added the assumption that the detach rate is small, that is, $k_{-1}^{(1)}\approx k_{-1}^{(2)}\approx 0$, for simplicity. 
Computation details are included in the appendix A.

\subsection{Multiple CAR T-cells bound to one glioma cell  model}

Following the same idea we have for the two-binding model, we can further build up a similar model for $n$ numbers of CAR T-cell binding to one glioma cell. We denote $I_j$ as the conjugate of $j$ CAR T-cells and one glioma cell. The notation $k_i^{(n)}$ denotes the rate of different reactions, where the superscript $(n)$ denotes the $n$-binding and the subscript $i$ denotes the $i$-th reaction in $n$-binding reaction. The governing equations of the $n$-binding model that includes the conjugates $I_1$, ..., $I_n$ can be written as follows. 
%

\begin{eqnarray}
    \label{eq:nbindslow} 
    \dot{T} &=& pT\frac{C}{g+C}-\theta T-k_1^{(1)}CT +\sum_{j=1}^n \alpha_j I_j  -\sum_{j=1}^{n-1}k_1^{(j+1)} I_j T \nonumber \\
    \dot{C} &=& \rho C(1-\frac{C}{K})-k_1^{(1)}CT+k_{-1}^{(1)}I_1+\sum_{j=1}^n \beta_j I_j \hspace{1.6cm} \nonumber \\
    \dot{I_1} &=& k_1^{(1)}CT-(\gamma_1+k_1^{(2)}T)I_1 +k_{-1}^{(2)}I_2 \hspace{2.75cm} \nonumber \\
    \dot{I_j} &=& k_1^{(j)}I_{j-1}T-(\gamma_j+k_1^{(j+1)}T)I_j + k_{-1}^{(j+1)} I_{j+1} \hspace{1.55cm} \nonumber \\
    \dot{I_n} &=& k_1^{(n)}I_{n-1}T- \gamma_n I_n   \hspace{5.0cm}
\end{eqnarray}
where $\alpha_j=\sum_{i=1}^j ik_{i+2}^{(j)} +k_{-1}^{(j)}$, 
$\beta_j=\sum_{i=1}^jk_{i+1}^{(j)}$, and 
$\gamma_j=\sum_{i=2}^{j+2}k_i^{(j)} + k_{-1}^{(j)}$.
The parameters and their biological meanings are summarized in Table \ref{Tbl:param}. 

In the following sections, we will test various hypotheses on the reaction rates to reproduce the phenomena of saturation of CAR T-cell therapy efficacy when the antigen receptor density of cancer increases. 

\begin{table} 
	\begin{tabular}{|c|l|} \hline 
	parameter & biological meaning  \\ \hline 
	$ \rho $  &  proliferation rate of glioma cells  \\ 
	$ K $  &  carrying capacity of glioma cells \\ 
	$ p $  & rate of CAR T-cell expansion induced by glioma  \\ 
	$ \theta $  & death rate of  CAR T-cells  \\ 
	$ g $  & steepness coefficient of   CAR T-cell recruitment  \\ 
	$k^{(n)}_1 $ & binding rate of CAR T-cell and  conjugate $I_{n-1}$   \\
	$k^{(n)}_{-1} $ & detaching rate of CAR T-cell and conjugate $I_{n}$  \\
	$k^{(n)}_{n+2} $ & death rate of glioma cells from conjugate $I_{n}$ \\
	$k^{(n)}_{i} |_{i=2}^{n+1}$ & death rate of CAR T-cells from conjugate $I_{n}$   \\
  \hline 
	\end{tabular}
	\caption{Model parameters and their biological interpretation}
	\label{Tbl:param}
\end{table}


\section{Stability analysis}
\label{sec:analysis}
In order to better understand the dynamics of CAR T-cell therapy with the slow and fast reaction models, we present  stability analysis of the one-to-one binding model.

\subsection{One-to-one binding slow reaction model  }

In this section, we study the stability of one-binding slow reaction model \eqref{eq:1bindslow}. There are four steady states $(T,C, I_1)$ in the slow reaction model 
$$(0,0,0), \quad (0,K,0),\quad (T_1,C_1,\frac{k}{\gamma} C_1T_1),\quad (T_2,C_2,\frac{k}{\gamma} C_2T_2),$$
where 
$$T_{1,2}=\frac{\rho(1-\frac{C_{1,2}}{K})}{k-\frac{\beta k}{\gamma}},$$
and 
\begin{eqnarray*}
C_{1,2}&=&\frac{(p-\theta-kg+\frac{\alpha kg}{\gamma})}{2(k-\frac{\alpha k}{\gamma})} \\ 
&\,& \pm \frac{ \sqrt{(p-\theta-kg+\frac{\alpha kg}{\gamma})^2-4(\frac{\alpha k}{\gamma}-k)(-\theta g)}}{2(k-\frac{\alpha k}{\gamma})}.
\end{eqnarray*}
Among these four states, we are interested in $(0,\displaystyle K,0)$ and $(T_1,C_1, \frac{k}{\gamma} C_1T_1)$, which represent CAR T-cell treatment failure and success respectively. We aim to find the parameter conditions that yield these two states, especially focusing on the CAR T-cell expansion rate induced by the presence of cancer, $p$.

\subsubsection{Escape from reaching the maximum size of tumor }

One of the steady states is $(0,\displaystyle K,0)$ that represents the tumor reaching its maximal capacity, or in other words, the CAR T-cell treatment failure. This steady state becomes stable if 
\begin{equation}
p<\left(\frac{g}{K}+1\right)\left(-\frac{K k\alpha}{\gamma}+\theta+Kk\right). 
\label{eq:p_cond1_slow}
\end{equation}
In other words, if the engineered CAR T-cell expansion rate $p$ is less than $(\frac{g}{K}+1)(-\frac{K k\alpha}{ \gamma }+\theta+Kk)$, the tumor reaches its maximum capacity. 
On the other hand, if the engineered CAR T-cells can proliferate enough so that $p\geq (\frac{g}{K}+1)(-\frac{K k\alpha}{ \gamma }+d+Kk)$, then the CAR T-cell therapy will prevent the cancer from growing to its maximum capacity. 
Therefore,  this condition provides the minimal level of CAR T-cell expansion rate to guarantee the escape from the scenario of reaching the maximal cancer size.

\subsubsection{Potential CAR T-cell treatment success }

Another steady state of our interest is the CAR T-cell treatment success state, 
$(T_1,C_1, \frac{k}{\gamma} C_1T_1)$, where we order the points as $C_1 < C_2$, so that $C_1$ is the state with a smaller cancer size. This state can appear when CAR T-cells are added. Then, the tumor is reduced to size $C_1$ compared to reaching its maximal capacity $K$. For this equilibrium point to exist, the following condition needs to be satisfied, 
\begin{equation}
\left( \sqrt{\theta} + \sqrt{g\left({\alpha k}/{\gamma}-k\right)}   \right)^2 \leq p . 
\label{eq:p_cond2_slow}
\end{equation}
This condition provides a minimum level of CAR T-cell expansion rate that makes it possible to succeed in CAR T-cell treatment, although the treatment success will depend on other parameters and initial CAR T-cell dosage. More detailed condition can be found in the appendix B. 


\subsection{One-to-one binding fast reaction model }

The stability analysis of the one-to-one binding fast reaction model Eq. \eqref{eq:Kuz} is comparable to the slow reaction model. The analysis can be found in many literature including \cite{Kuznetsov1994,Cho2020}. Here, we briefly summarize it to compare it with the slow reaction model. Identical to the slow reaction model, there are four possible steady states $(T,C)$, 
$$(0,0), \quad (0,K), \quad (T_1,C_1), \quad (T_2,C_2), $$
where 
\begin{eqnarray*}
T_{1,2}&=&\frac{\rho\left(1-\frac{C_{1,2}}{K}\right)}{n}, \\
C_{1,2}&=&\frac{(p-\theta-mg)\pm \sqrt{(p-d-mg)^2-4mg\theta}}{2m}.
\end{eqnarray*}
The steady state that the tumor growing to its maximum capacity, $(T,C)=(0,K)$, becomes stable if 
\begin{equation}
p<\left(mK+\theta \right)\left(\frac{g}{K}+1\right).  
\label{eq:p_cond1}
\end{equation}
This condition is comparable to Eq.  \eqref{eq:p_cond1_slow}, which provides the condition that the CAR T-cell expansion rate prevents the cancer from growing to its maximum. 
The other condition related to the equilibrium point $(C_1,T_1)$, the CAR T-cell therapy success case, is 
\begin{equation}
\left(\sqrt{\theta}+\sqrt{mg}\right)^2\leq p. 
\label{eq:p_cond2}
\end{equation}
The CAR T-cell needs to expand at least this level to have chances for successful CAR T-cell treatment. We remark that the other equilibrium points $(0,0)$ and $(T_2,C_2)$ are always saddles, so that they are not of our interest.

\section{Numerical study of glioma cells and CAR T-cells interaction }
\label{sec:simulation}

Our experimental data consist of multiple doses of CAR T-cells and multiple types of glioma cells based on the antigen receptor densities. The data includes glioma cells with low, medium, and high antigen receptor densities, where the latter is likely to have a better response to CAR T-cell treatment, although not strictly better. For each density level, the experiments were initialized with different mixture ratios between CAR T-cells and glioma cells as 1:5, 1:10, and 1:20. Using this rich dataset, we numerically study the proposed multiple CAR T-cell binding model to find a better model that describes the experimental data. First, we show that the slow reaction model Eq. \eqref{eq:1bindslow} describes the data more accurately compared to the fast reaction model Eq. \eqref{eq:Kuz}. Then, we compare different assumptions in the reaction rates to investigate the saturation of CAR T-cell treatment efficacy regarding the antigen density levels.

\subsection{Comparison between fast reaction and slow reaction models }

\begin{figure}[!b]
\centerline{ (a) \hspace{3.5cm} (b) }
\centerline{
    \includegraphics[width=4cm]{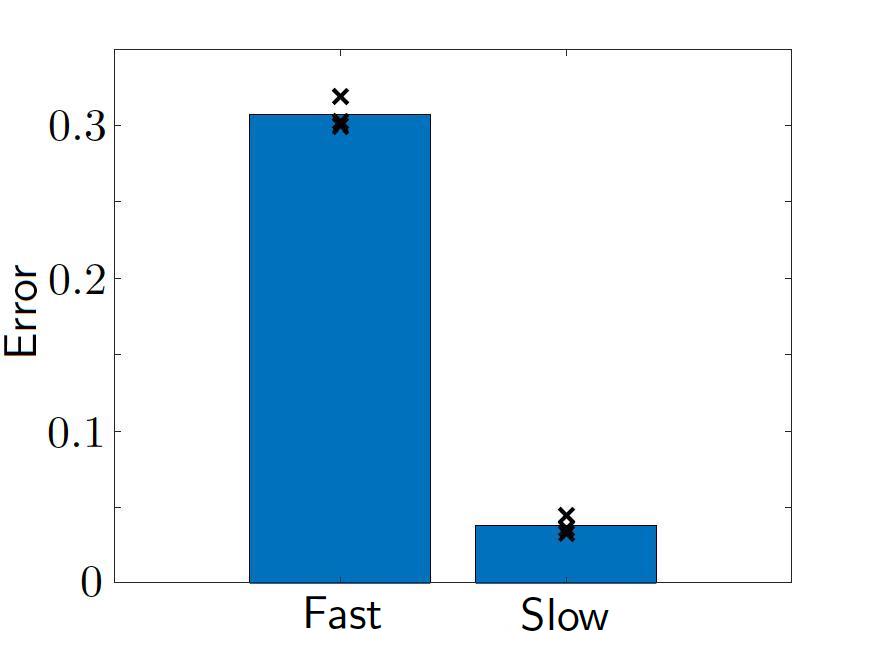}
    \includegraphics[width=4cm]{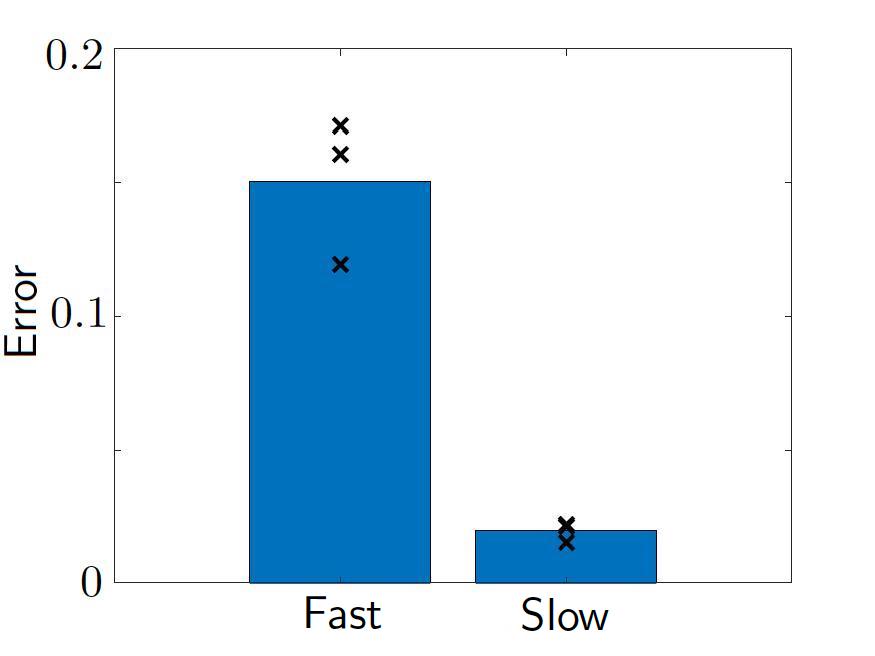}}
\centerline{
    \includegraphics[width=4cm]{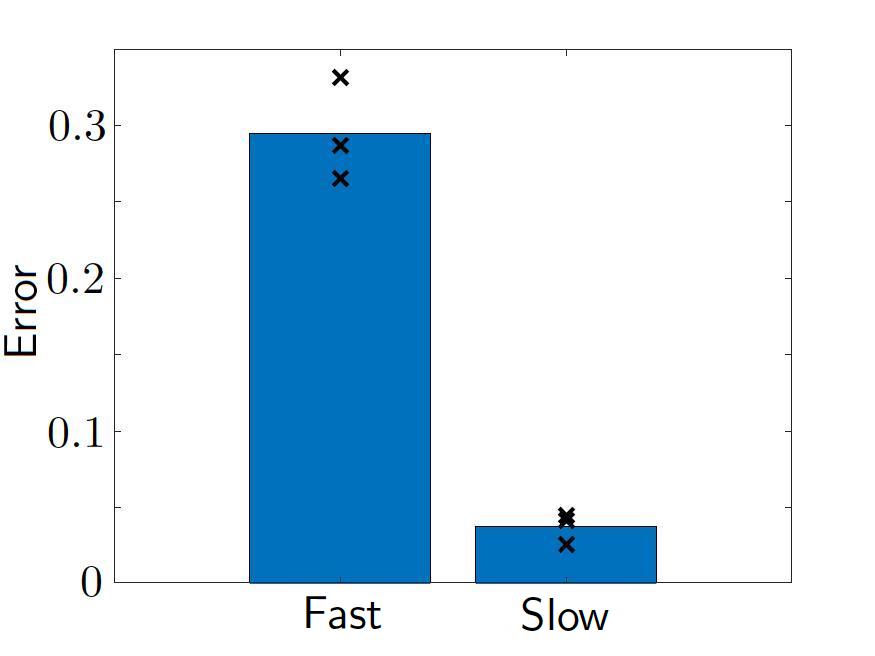}
    \includegraphics[width=4cm]{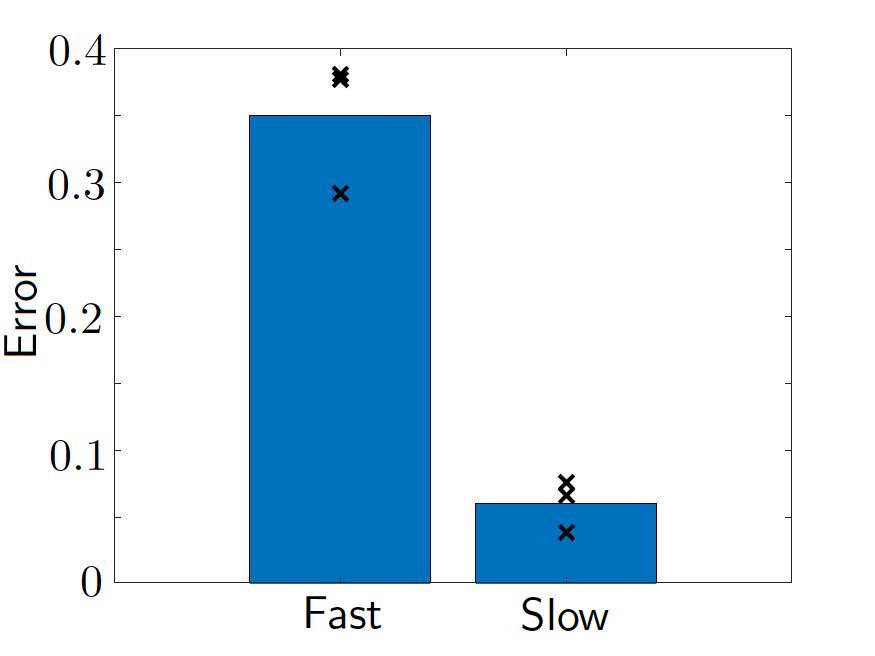}}
\centerline{
    \includegraphics[width=4cm]{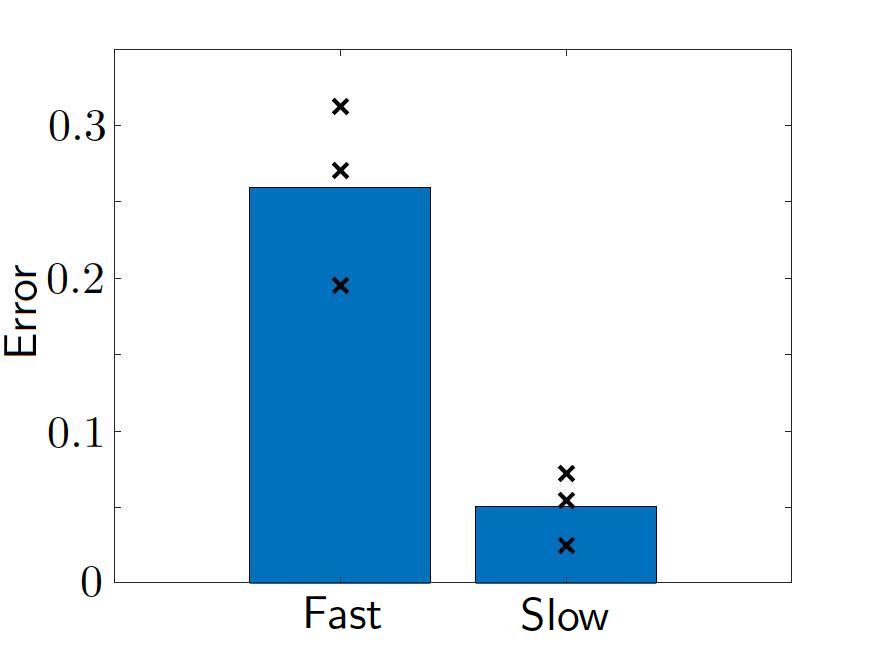}    \includegraphics[width=4cm]{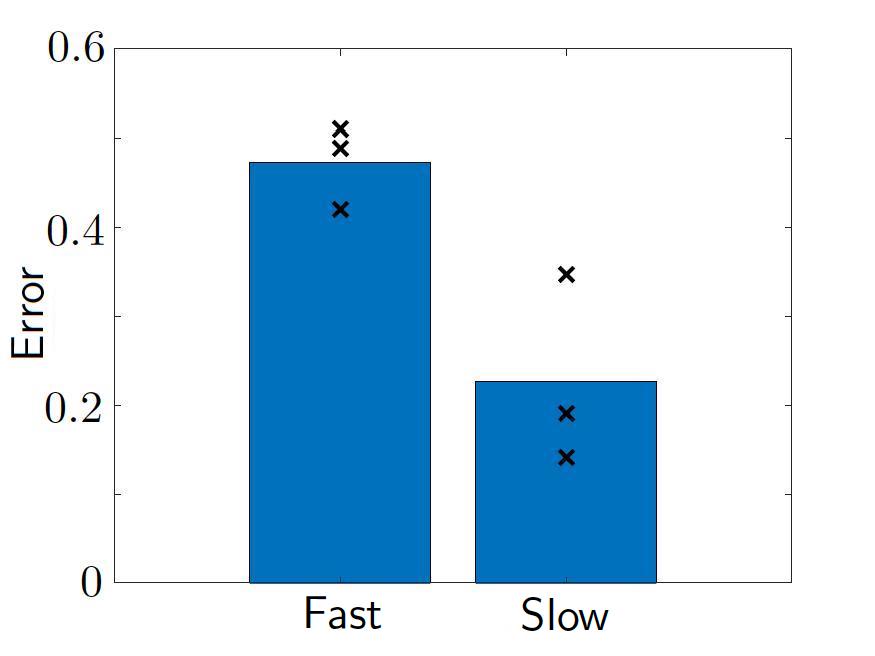}}
    \caption{Comparison of model fit errors between the fast reaction model Eq. \eqref{eq:Kuz} and the slow reaction model Eq. \eqref{eq:1bindslow}. The shown bar plots are the average of the three marked errors ($\times$) for fitting the data of low antigen receptor density glioma (a) and high antigen receptor density glioma (b), with CAR T-cell to cancer ratio 1:5 (top), 1:10 (middle), and 1:20 (bottom). It can be seen that slow reaction model has smaller errors in all cases.}
    \label{fig:fastVSslowError}
\end{figure}

The fast reaction model, for example, Eq. \eqref{eq:Kuz} does not describe the dynamics of the conjugates $I_j$, assuming that they reach the equilibrium state immediately. However, in reality the reaction is not always fast enough as it takes time for the CAR T-cells to detect and infiltrate the glioma cells, which is also depicted in our experimental data. Therefore, we expect the slow reaction model \eqref{eq:1bindslow} to describe the data more accurately. 
Our hypothesis is confirmed in Figure \ref{fig:fastVSslowError}, where we compare the accuracy of the slow reaction model Eq. \eqref{eq:1bindslow} and the fast reaction model Eq. \eqref{eq:Kuz}. The error is computed as the sum of squares between the data and the calibrated model fit. In these plots, the errors from the three sets of data with the same receptor density (low and high) and CAR T-cell ratio (1:5, 1:10, 1:20) are marked as black crosses, and the bar shows the average of the errors. We observe that the slow reaction model results in significantly smaller errors for all cases. Let us comment on each data set more closely.

\begin{figure}[!b]
\centerline{ Fast \hspace{3.5cm} Slow }
\centerline{
    \includegraphics[width=4.2cm]{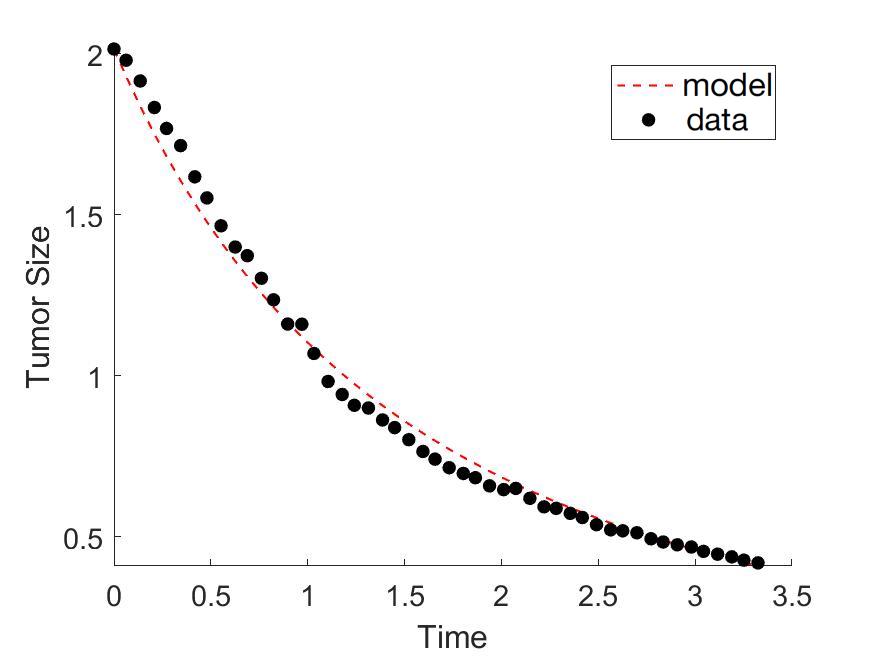}
    \includegraphics[width=4.2cm]{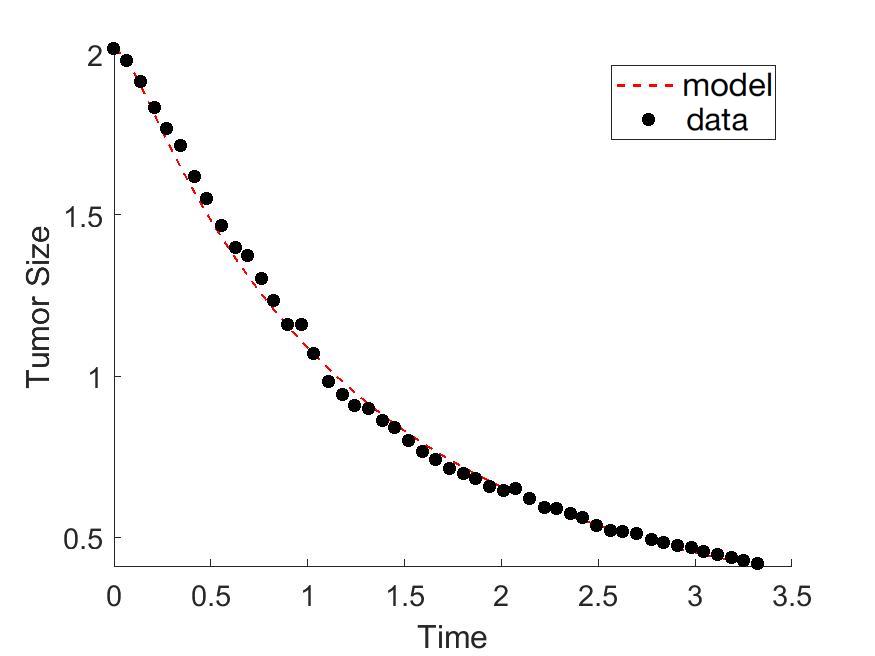}
    }
\centerline{
    \includegraphics[width=4.2cm]{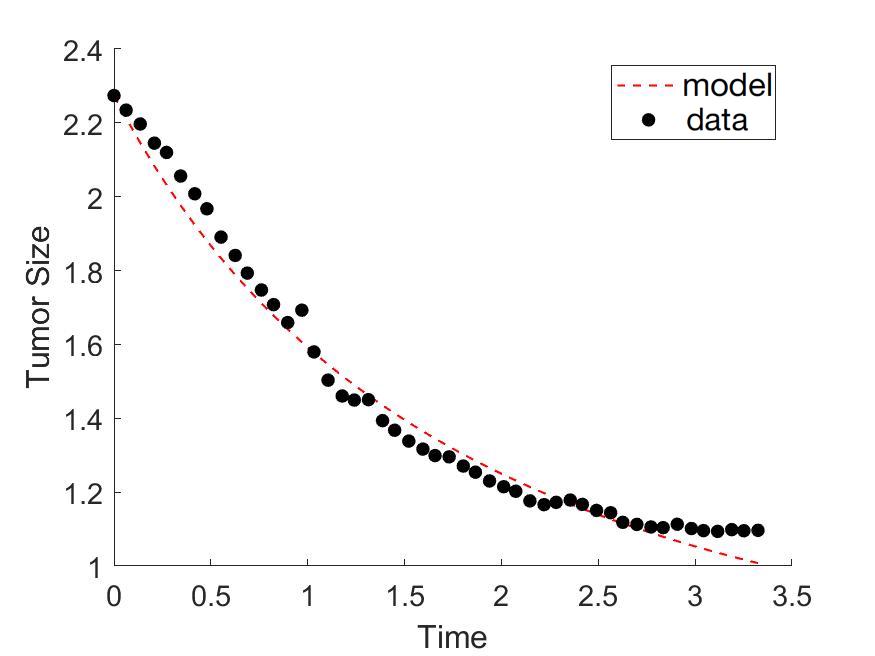}
    \includegraphics[width=4.2cm]{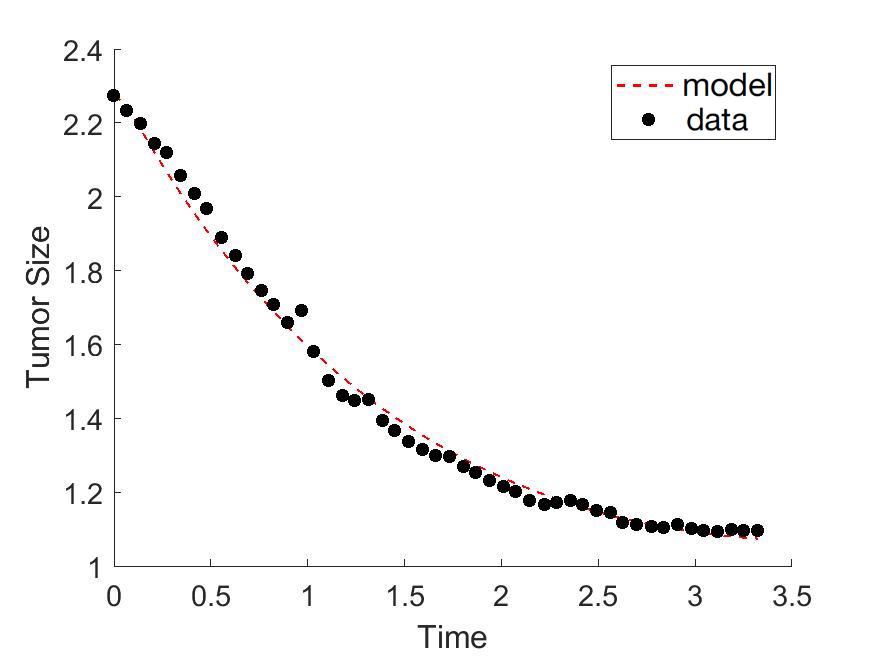}
}
\centerline{
    \includegraphics[width=4.2cm]{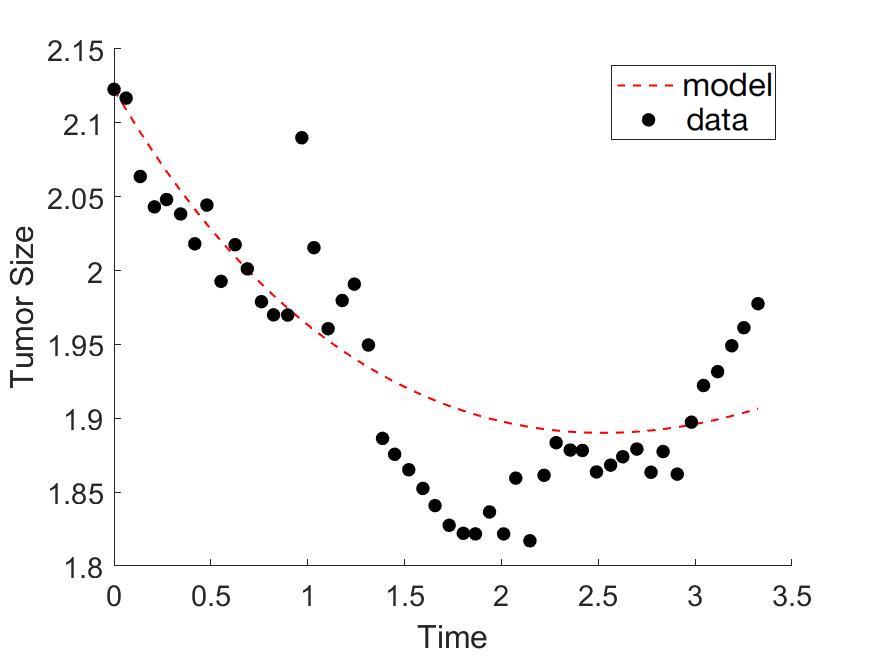}
    \includegraphics[width=4.2cm]{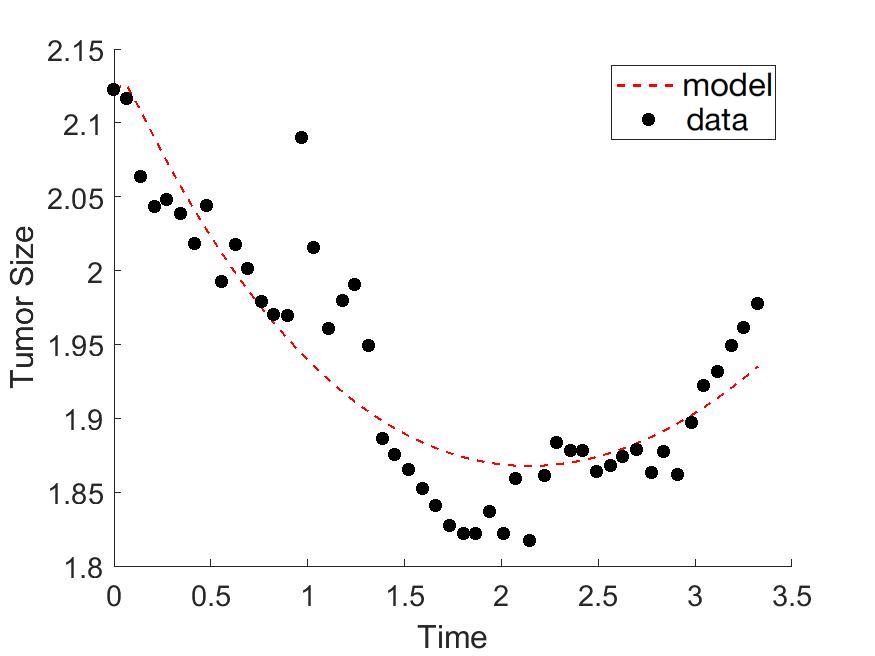}
}
\caption{\textbf{Low IL13R$\alpha$2 antigen density glioma} Comparison of the calibrated model fits between the fast reaction model Eq. \eqref{eq:Kuz} (Fast) and the slow reaction model Eq. \eqref{eq:1bindslow} (Slow). The shown results are for low receptor density cancer with CAR T-cell to cancer ratio 1:5 (top), 1:10 (middle), and 1:20 (bottom). The calibrated model fits using the slow reaction model are closer to the data points, demonstrating that the slow reaction model more accurately describes the experimental data than the fast reaction model for the case of low receptor density.}
    \label{fig:fastVSslowL}
\end{figure}

Figure \ref{fig:fastVSslowL} compares the accuracy of slow binding and fast reaction models calibrated to the {\em low} receptor density glioma data. We observe that the slow reaction model does have a better fitting than the fast reaction model. The top row shows the case of initial ratio 1:5, where we can see that the data points are closer to the calibrated model curve especially near $t=0.5$ and $t=1.5$ using the slow reaction model. Similarly in the second row that corresponds to the 1:10 ratio, the fitting of the slow reaction model is significantly better at capturing the saturating tail at the later time points after $t=3$. The last row shows the result of initial ratio 1:20, which is the case with the smallest CAR T-cell dosage. Again, the slow reaction model better describes the non-monotonic data.  In all three cases of low antigen receptor density glioma experiments, we observe that the slow reaction model is more accurate, especially when the concavity is nonzero. The reason behind this result is that the reaction speed is  slower when the antigen receptor density is low, which makes the slow reaction model more appropriate than the fast reaction model. 

\begin{figure}[!b]
\centerline{ Fast \hspace{3.5cm} Slow}
\centerline{
    \includegraphics[width=4.2cm]{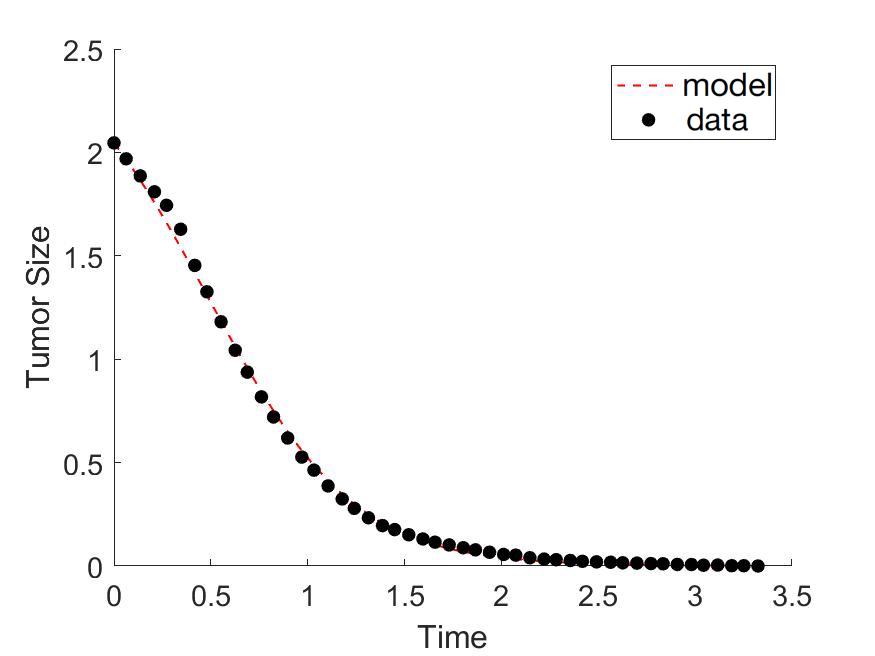}
    \includegraphics[width=4.2cm]{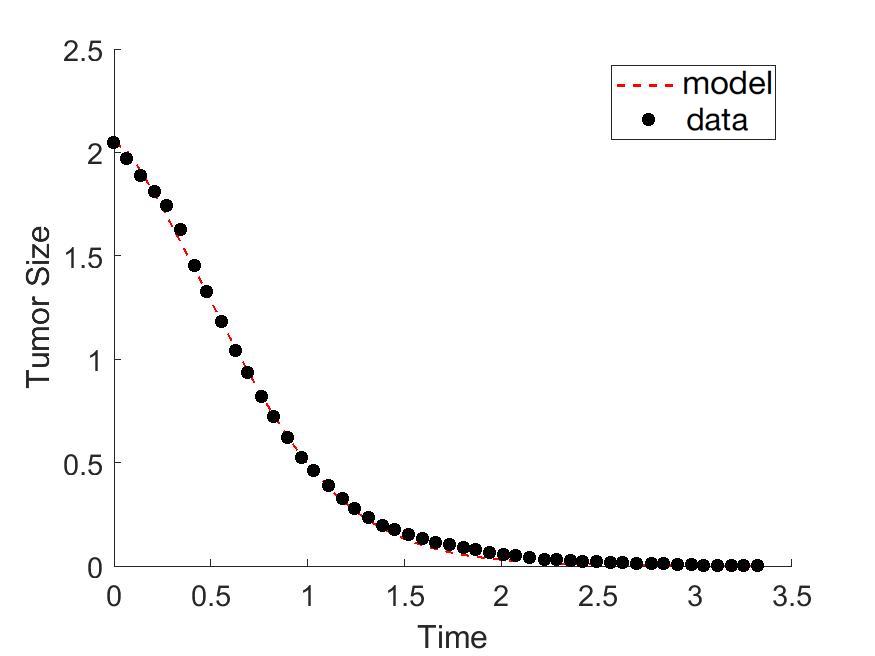}}
\centerline{
    \includegraphics[width=4.2cm]{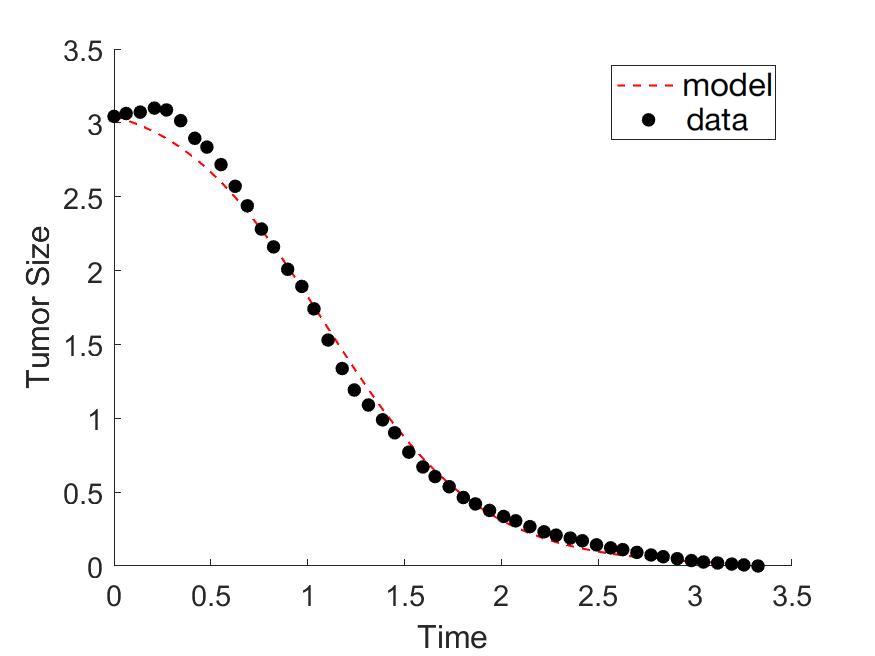}
    \includegraphics[width=4.2cm]{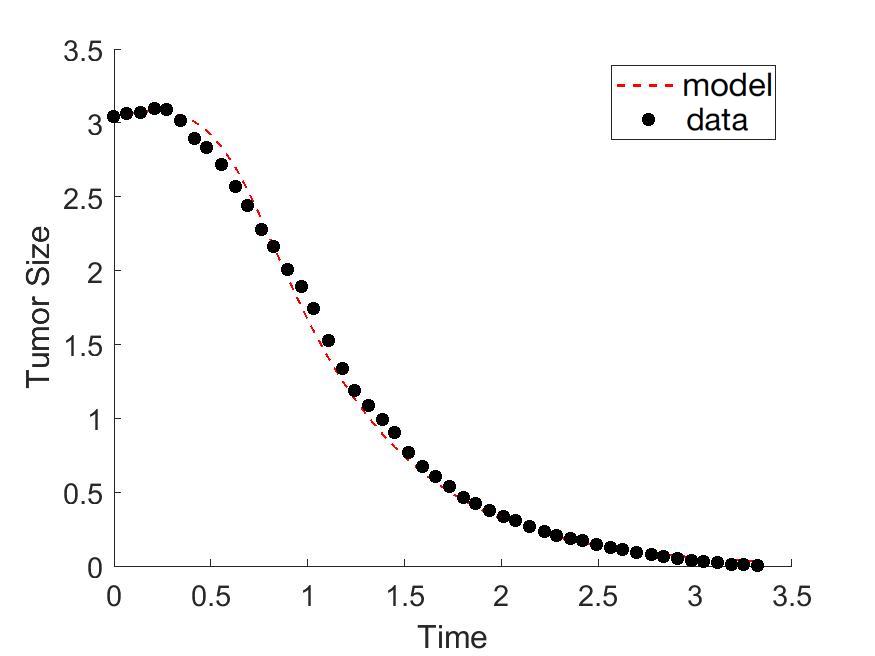}}
    \caption{\textbf{High antigen receptor density glioma} Comparison of calibrated model between the fast reaction model Eq. \eqref{eq:Kuz} (Fast) and the slow reaction model Eq. \eqref{eq:1bindslow} (Slow). The shown results are for high receptor density cancer with CAR T-cell to cancer ratio 1:5 (top), 1:20 (bottom). The calibrated model fits between the two models are similarly good in the case of 1:5 ratio, that is the case of higher dosage. Since the high receptor density glioma cells with high dosage of CAR T-cells is the case that the reaction can be the most efficient, the fast reaction model is not significantly better when it comes to this case.}
    \label{fig:fastVSslowH}
\end{figure}

On the other hand, we hypothesize that the fast reaction model could be more appropriate to fit the experiments with {\em high} antigen receptor density glioma. Since the high receptor density glioma cells have more receptors for the CAR T-cells to bind, the reactions can be more efficient. We observe that from the data plotted in Figure \ref{fig:fastVSslowH} that the tumor size decay faster compared to the low receptor density case. This is the case especially when the glioma cells are mixed with a large number of CAR T-cells, that is, the 1:5 ratio case among our experiments. The top row of Figure \ref{fig:fastVSslowH} shows the 1:5 ratio mixture, where the CAR T-cells eliminate the glioma cells most rapidly. The model fits using the slow and fast reaction models both look accurate, and we confirm that this is the case that the fast reaction model can accurately capture the data. No big difference can be seen between the slow and fast reaction model fits, although the error is smaller using the slow reaction model (see Figure \ref{fig:fastVSslowError}(b)). 
Nonetheless, in case of lower dosages of CAR T-cells, for instance, 1:20 mixture, the slow reaction model again becomes visibly more accurate, being able to capture the initial bump of the data.

\begin{figure}[!htb]
\centerline{\hspace{.2cm} Fast}
\centerline{ 
    \includegraphics[width=4.4cm]{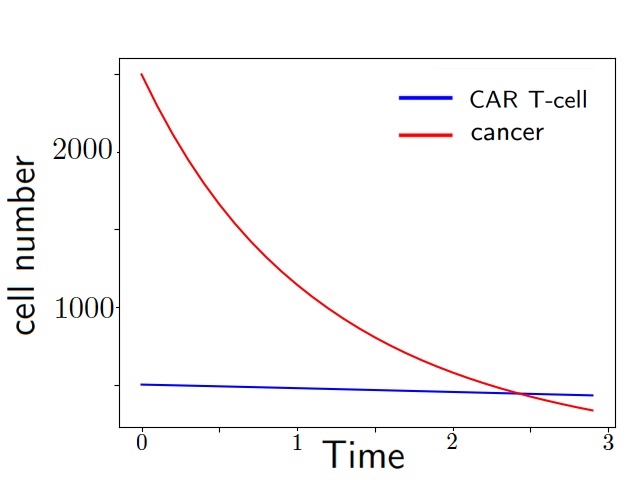}
    }
\centerline{\hspace{.2cm} Slow}
\centerline{ 
    \includegraphics[width=4.2cm]{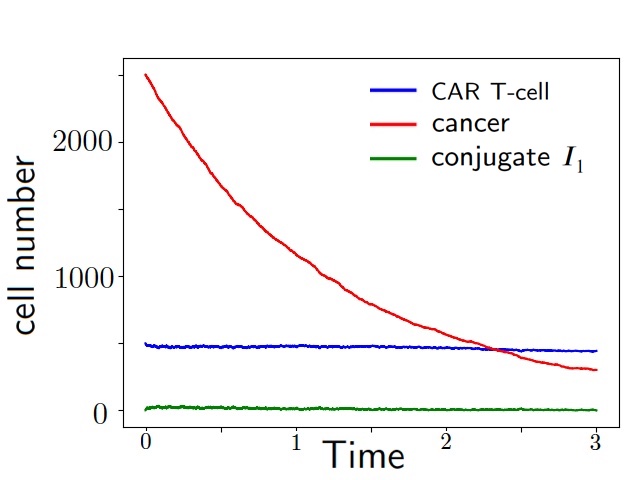}
    \includegraphics[width=4.2cm]{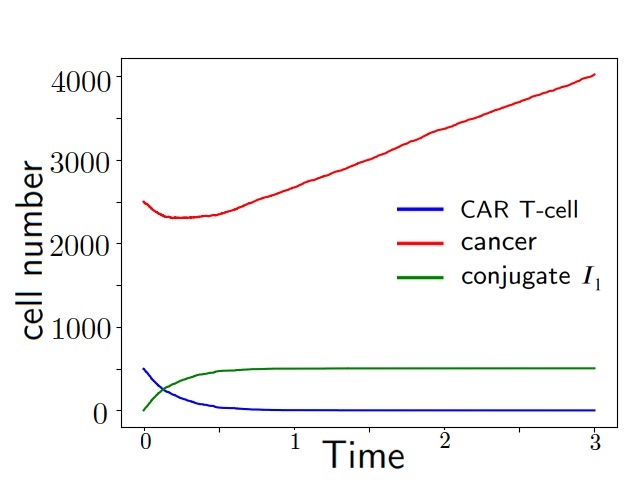}
    }
    \caption{
    Dynamics of cancer, CAR T-cell, and conjugate $I_1$ using two parameter sets of slow reaction model (bottom) that reduces to the same fast reaction model (top) parameter set. When the cancer killing rate is large as $k_3^{(1)}=100$ (bottom, left), the dynamics of slow and fast reaction models agree. However, when the cancer killing rate is small as $k_3^{(1)}=0.01$ (bottom, right), the conjugate $I_1$  have non-trivial dynamics and the dynamics of slow reaction model differs from the fast reaction model.  
    }
    \label{fig:Stochastic1}
\end{figure}

We conclude that if the reaction between the glioma cells and CAR T-cells is not immediate, the slow reaction model \eqref{eq:1bindslow} more accurately describes their interaction. This happens especially when the glioma cells have low antigen receptor density and CAR T-cell numbers are small. 
%
In addition to the model fit comparison, we argue that the slow reaction model can either match the fast reaction model or deviate from it based on the parameter values. We remark that the slow reaction model has four interaction parameters that defines two parameters of fast reaction model, $n$ and $m$, presented in Eq. \eqref{eq:param_relation}. 
Therefore there are multiple sets of parameters of slow reaction model that reduces to an identical fast reaction model. Figure \ref{fig:Stochastic1} shows such an example. The bottom figures are computed from two distinct parameter sets of the slow reaction model that yields identical $n$ and $m$ values, $n=2.785$ and $m=0.0223$, for the fast reaction model. 
Large values of $k_3^{(1)}$ indicate fast reaction, for example, when $k_3^{(1)}=100$, the $I_1$ conjugate dynamics are trivial and the slow reaction model agrees the fast reaction model. However, when $k_3^{(1)}=0.01$, we observe the $I_1$ conjugates number increasing and the slow reaction model show distinct result from the fast reaction model. While the fast reaction model shows the number of glioma cells declining, the slow reaction model with a different choice of $k_3^{(1)}$ changes the glioma cell dynamics from decreasing to increasing. This example shows the richer dynamics that the slow reaction model contains. 

\subsection{Simulation of multiple-binding slow reaction models}

\begin{figure}[!b]
    \centerline{\hspace{.6cm}(a) \hspace{3.34cm} (b)}
    \centerline{
    \includegraphics[width=8.0cm]{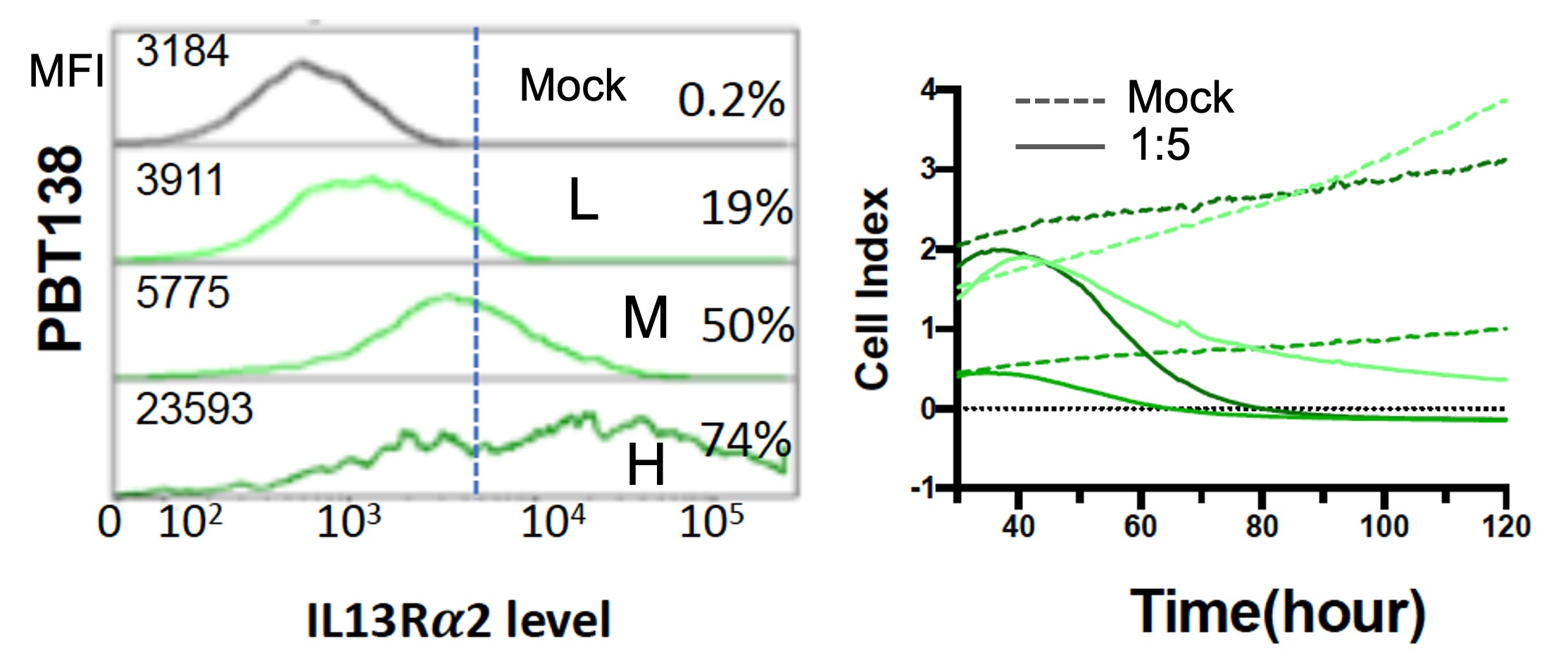}
    }
    \centerline{\hspace{0.15cm}(c) \hspace{3.55cm} (d)}
    \centerline{
    \includegraphics[width=4.0cm]{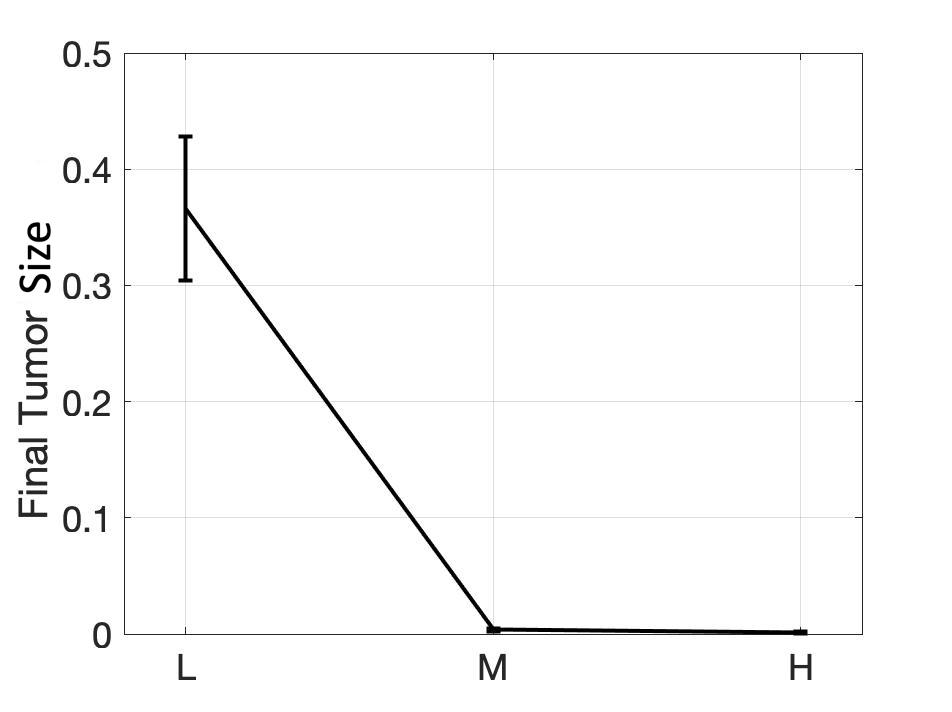}
    \includegraphics[width=4.0cm]{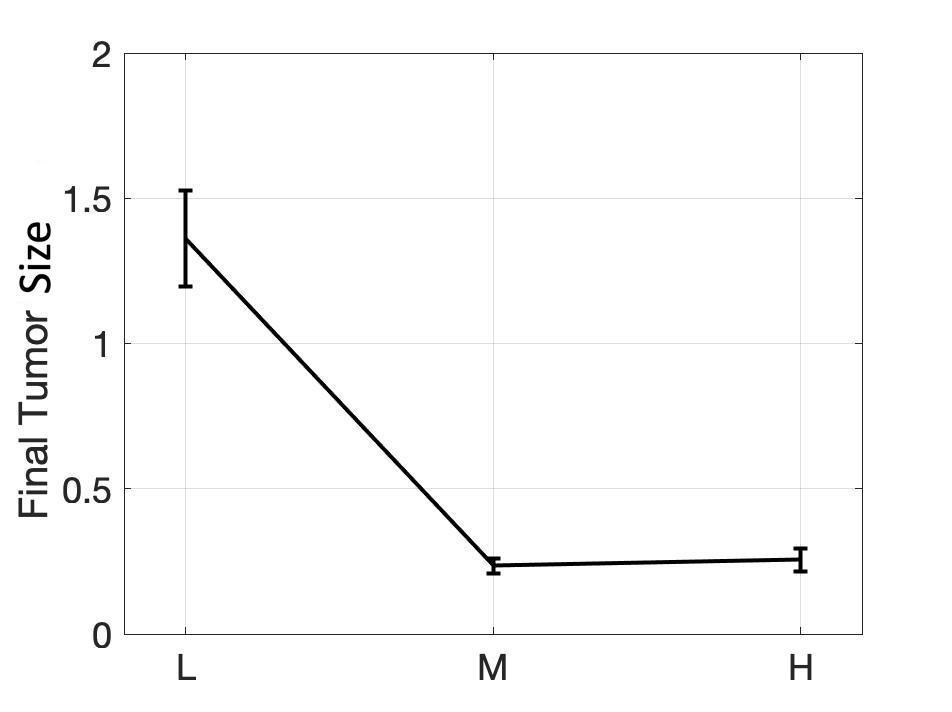}
    }
    \caption{Antigen expression measured with flow cytometry (mean florescence intensity MFI, and the percentage of cells positive) for cell line PBT138  (mock, low (L), medium (M), high (H)) reported in Sahoo et al. \cite{Sahoo2019} (a). Dynamics of the size of glioma cell population measured by xCELLigence (b). Glioma tumor size data (CI) after mixed with CAR T-cells in 1:5 ratio (c) and 1:10 ratio (d) in terms of antigen receptor density level of glioma cells: low (L), medium (M), and high (H) are shown as well.  }
    \label{fig:FinalTumVol_data}
\end{figure}

Having compared the slow and fast reaction models in the previous section, we now study the multiple CAR T-cells binding model Eq. \eqref{eq:nbindslow}. In particular, we focus on the experimental results of glioma cells with different levels of antigen receptor density: low, medium, and high. When mixed with the same number of CAR T-cells, glioma cells with low antigen receptor density respond less than the glioma cells with medium antigen receptor density. In other words, the decay rate of glioma cell is positively correlated with the antigen receptor density in general. However, an interesting observation was made in \cite{Sahoo2019} that the effectiveness CAR T-cell therapy saturated after a certain receptor density level. As shown in Figure \ref{fig:FinalTumVol_data}, the high receptor density cells did not respond better than the medium receptor density glioma cells, or responded rather worse. We aim to study this phenomena using the multiple CAR T-cells binding models Eq. \eqref{eq:nbindslow}. Assuming that the antigen receptor density levels of glioma cells determines the number of CAR T-cells that can bind to a single glioma cell, we associate the one CAR T-cell binding model ($n=1$) to the low density receptor glioma, and multiple CAR T-cells binding model up to the five binding ($n=5$) to higher density receptor levels of glioma. 
One problem we face for the multiple CAR T-cells binding models is that the number of interaction parameters $k_i^{(j)}$ increases as the number of binding increases. Since we do not have data for the dynamics of conjugates $I_j$, it is difficult to estimate parameters from the data. Therefore, using the parameter set estimated for the one-binding model, we test the following two different hypotheses describing the relationships between the reaction rates for the multiple binding models with more than one binding. 
\begin{itemize}
    \item Hypothesis 1: Assume that the reaction rates are {\em uniform} across conjugates $I_j$. 
    
     $\bullet$ The CAR T-cell attaches to the glioma cell or the conjugate $I_j$ with an identical rate, i.e.,
     \begin{equation}
         k_1^{(1)}=k_1^{(j)}, \quad \textrm{ for all } j = 2, 3, ..., n. 
         \label{eq:k_condition1}
    \end{equation}
         
     $\bullet$ The reactions that the glioma cell dies from the conjugate $I_j$ have the same rates across number of bindings, i.e., 
     \begin{equation}
        k_{1+2}^{(1)}=k_{i+2}^{(j)}, \quad \textrm{ for all } j = 2, 3, ..., n.         
        \label{eq:k_condition2}
    \end{equation}

     $\bullet$ The reactions that glioma cells' survive also have the same reaction rates with equal chances of CAR T-cells dying i.e, 
     \begin{equation}
        k^{(n)}_{j}=\frac{ {n \choose j-1}}{2n^2-6n+7}k_2^{(1)}, \,\, \text{ for } j=2,3,...,n+1, 
        \label{eq:k_condition3}
    \end{equation}
    so that $k^{(1)}_2=\sum\limits_{i=2}^{n+1}k_i^{(n)}$.

    \vspace{5mm}
    \item Hypothesis 2: Instead of assuming that the reaction rates are uniform across the conjugates $I_j$, in this hypothesis, we assume that the reaction rates are {\em non-uniform}, either increasing or decreasing, and saturate after a certain number of CAR T-cells bind. 
    
    $\bullet$ The reaction rates of a new CAR T-cell attaching to the conjugate $I_j$ decrease geometrically, i.e.,
     \begin{equation}
         k_1^{(j)}=\frac{k_1^{(1)}}{M^{j-1}},  \quad\textrm{ for all } j = 2, 3, ..., n. 
         \label{eq:k_condition1-3}
    \end{equation}
    for some positive integer $M\geq 1$. 
    
    $\bullet$ The glioma cells' death rate increases geometrically  as the number of bindings increases, but saturates after three CAR T-cells binding, i.e.,
     \begin{eqnarray}
   &\,& k^{(2)}_4=k^{(1)}_3 L,\,\,\,  k^{(3)}_5=k^{(1)}_3 L^2, \nonumber \\ 
   &\,& k^{(j)}_{j+2}=k^{(3)}_5,  \quad\textrm{ for all } j = 4, 5, ..., n. \hspace{1.5cm}
         \label{eq:k_condition2-3}
    \end{eqnarray}
    for some positive integer $L\geq 1$. \\
    The reactions that glioma cells' survive are distributed as in hypothesis 1, Eq. \eqref{eq:k_condition3}. 

\end{itemize} 

\begin{figure}[!htb]
    \centerline{Low \hspace{4cm} High}
    \centerline{
    \includegraphics[width=4.2cm]{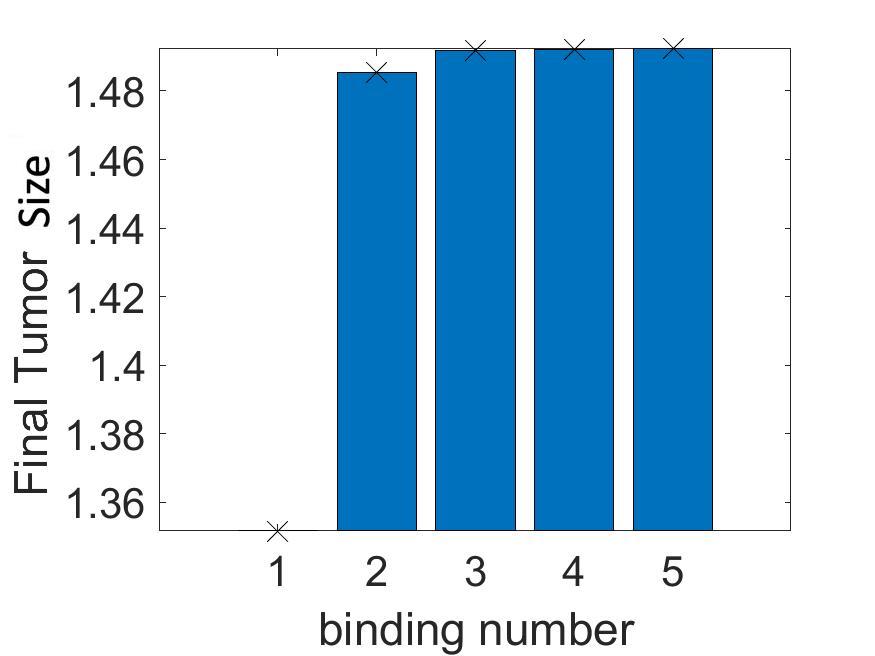}
    \includegraphics[width=4.2cm]{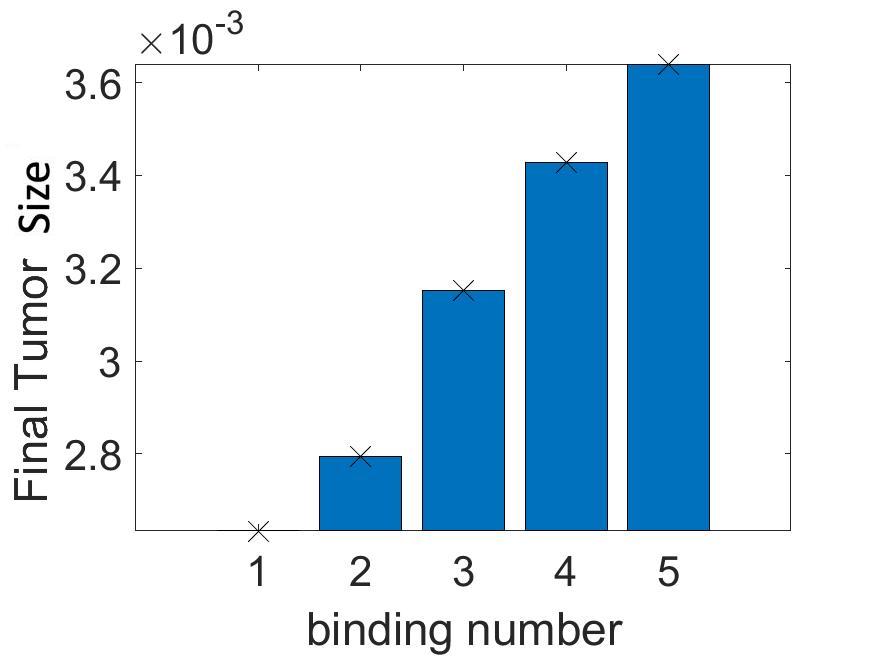}
    
    }
    \centerline{
    \includegraphics[width=4.2cm]{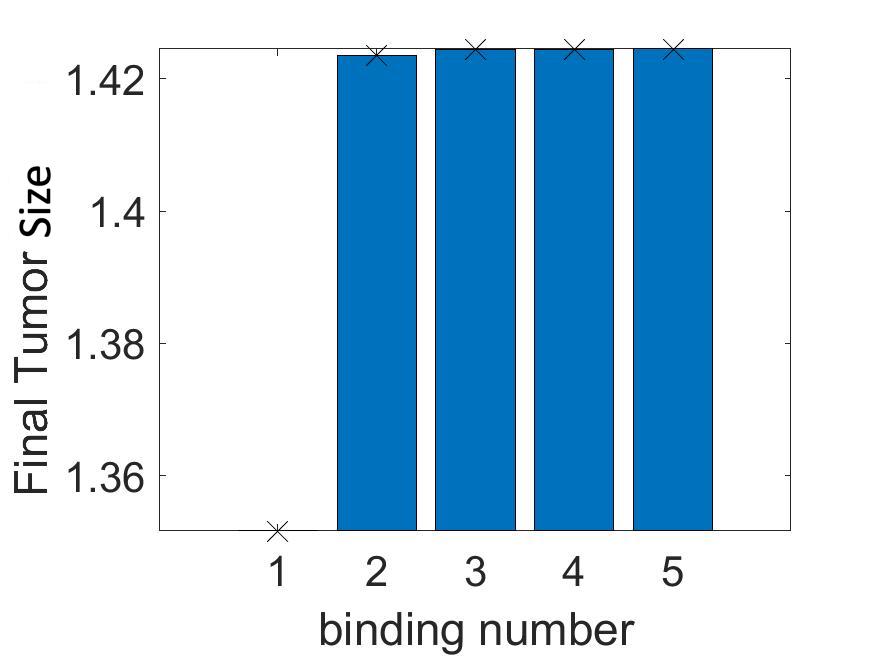}
    \includegraphics[width=4.2cm]{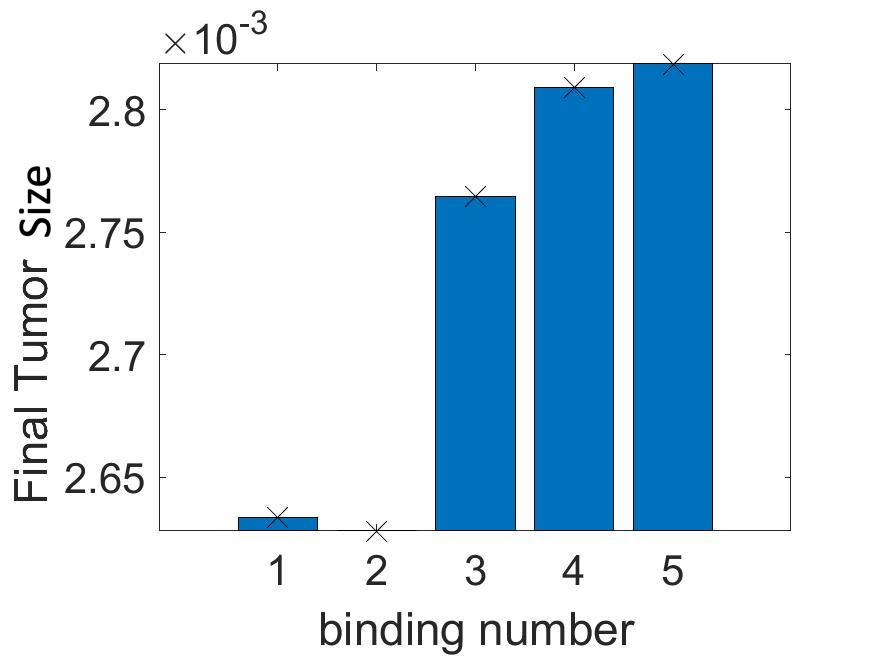}
    }
    \caption{{\em Hypothesis 1.} Final tumor size (CI) using the $n$-binding model Eq. \eqref{eq:nbindslow} assuming up to $n$ number of CAR T-cells binding to glioma cell forming the conjugates $I_1$, ..., $I_n$. We consider $n = 1,2,...,5$. 
    The kinetic rate parameters of the multiple CAR T-cells binding model  are taken by hypothesis 1 (top) and hypothesis 2 with $M=2,\, L=1$ (bottom). 
    In fact, the tumor size does not decay, but rather increases as the number of binding $n$ increases, in both low and high antigen receptor density cases.  }
    \label{fig:FinalTumVol_Try1}
\end{figure}

We compare the two hypotheses to find which assumption yields the experimental data we have considering different levels of antigen receptor densities in glioma cells. In particular, we compare the final tumor size after 3 days of CAR T-cell treatment, i.e. $C(t)$ at $t=3$. 
In the top row of Figure \ref{fig:FinalTumVol_Try1}, we show the results testing hypothesis 1. The results do not match what we observe in the experiments, as the final tumor size increases from one-binding to five-binding models. We presume that this is due to the relative reaction rate of cancer death decreasing as the number of reaction increases. 
Hence, by simply considering uniform reaction rates for one to multiple CAR T-cell conjugates as in hypothesis 1, we cannot recover the dynamics of glioma cells responding better in higher antigen receptor density levels. 

\begin{figure}[!htb]
    \centerline{Low \hspace{4cm} High}
    \centerline{
    \includegraphics[width=4.2cm]{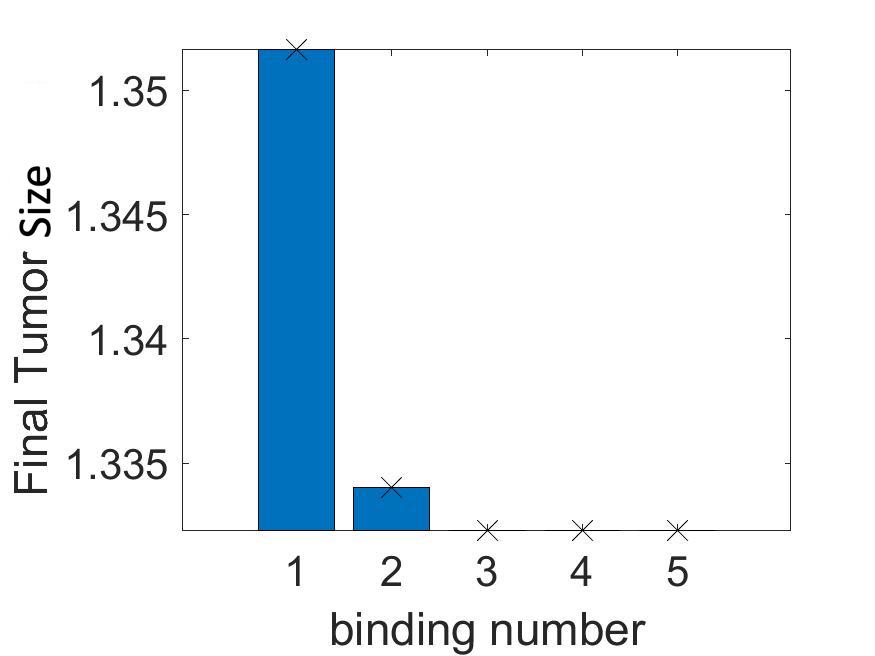}
    \includegraphics[width=4.2cm]{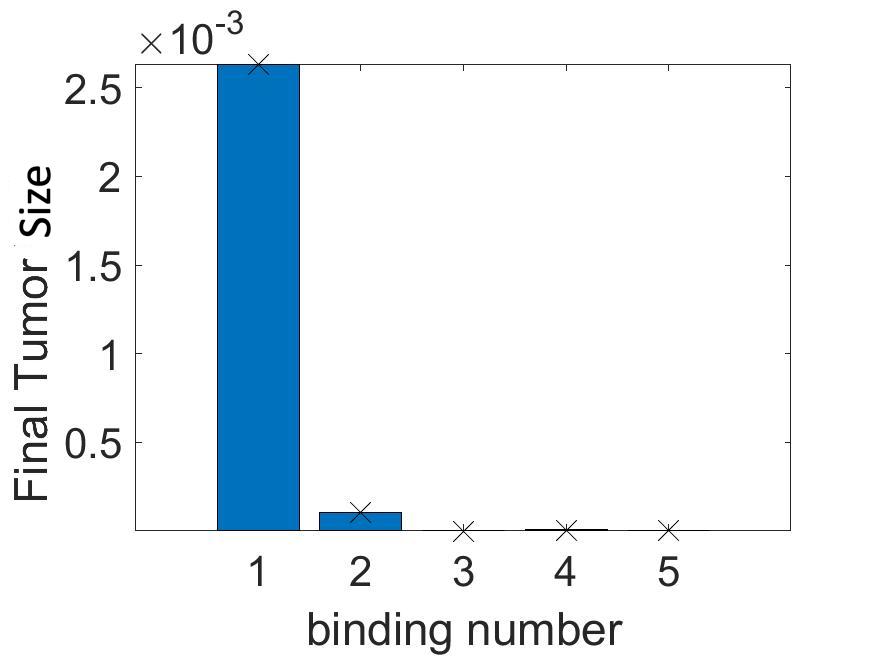}}
    \centerline{
    \includegraphics[width=4.2cm]{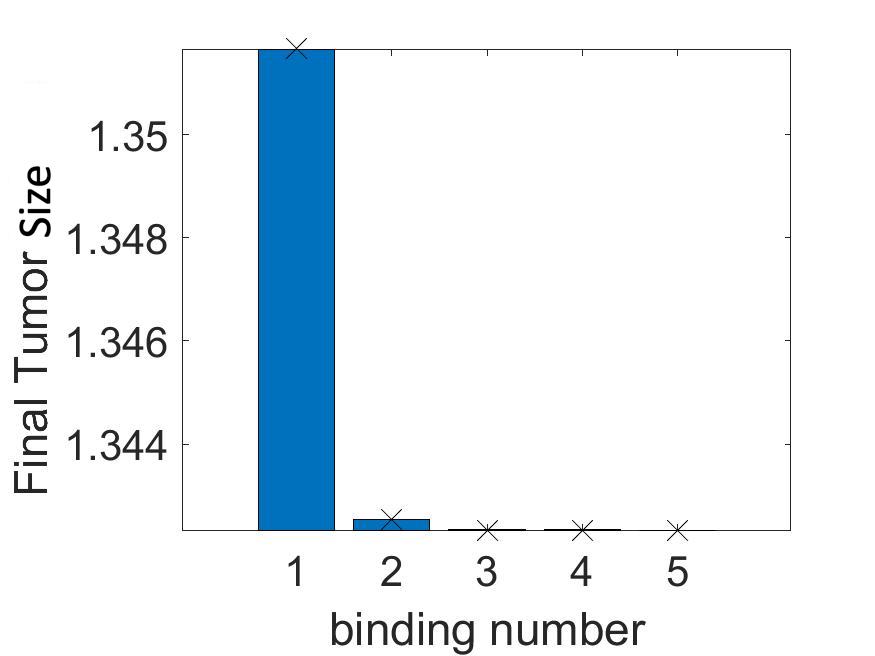} 
    \includegraphics[width=4.2cm]{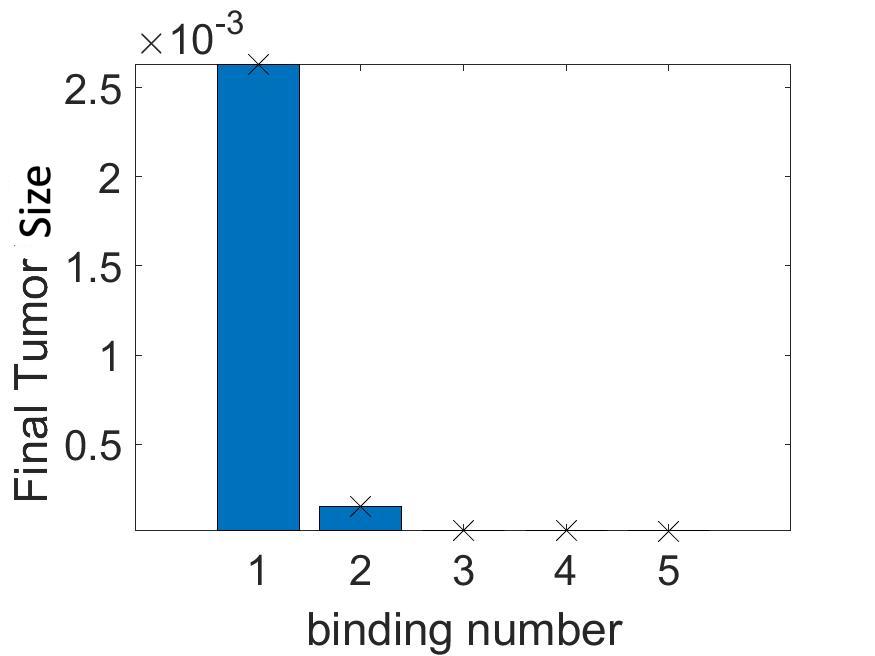}
    }
    \caption{{\em Hypothesis 2.} Final tumor size (CI) using the $n$-binding model Eq. \eqref{eq:nbindslow} assuming up to $n$ number of CAR T-cells binding to glioma cell forming the conjugates $I_1$, ..., $I_n$. We consider $n = 1,2,...,5$. 
    The kinetic rate parameters of the multiple CAR T-cells binding model are taken by hypothesis 2 with $M=1,\, L=2$ (top) and $M=2,\, L=2$ (bottom). 
    Unlike what we had in hypothesis 1, there is no increasing from two-bindings to three-bindings for all densities, and all results show saturation of final tumor size after three-bindings. Thus the relationships between reaction rates that we consider in hypothesis 2 with $L> 1$ describes the experimental data regarding the cancer antigen  density receptor level.}
    \label{fig:FinalTumVol_Try3}
\end{figure}

Nevertheless, results from hypothesis 2 show distinctive outcomes from hypothesis 1. The simulation results taking different values of $M$ and $L$ are shown in the bottom row of Figure \ref{fig:FinalTumVol_Try1} and Figure \ref{fig:FinalTumVol_Try3}. In particular, Figure \ref{fig:FinalTumVol_Try3} shows the result of  $M=1,\,L=2$, and $M=2,\,L=2$. It turns out that we no longer have the increase of tumor size as $n$ increases, instead we have a decrease for both low and high densities. We can further observe that the reaction saturates after the numbers of binding become more than three, as the final tumor size remains very small eventually. However, if we consider $M=2,\,L=1$, the tumor size increases a $n$ increases, and the results again deviate from the experimental data as shown in the bottom row of Figure \ref{fig:FinalTumVol_Try1}. 
Thus, we conclude that the hypothesis 2 on the reaction rates with $L> 1$ describes the experimental data regarding the decay and saturation of tumor size with respect to the antigen density receptor level. It makes sense that $L$ is the critical parameter for this result, since it is the parameter that makes the death rate of the glioma cells increase as $n$ increases.

\section{Summary and future work}
\label{sec:conclution} 

Here we developed an ODE model extending \cite{Kuznetsov1994}. While the original model considers one conjugate $I_1$ of one glioma cell and one CAR T-cell, and assumes that the dynamics of $I_1$ conjugate is in equilibrium, our model considers multiple conjugates $I_j$, $j=1,...,n$ with more than one CAR T-cells binding to the glioma cell, and follow their dynamics. We denote the original model as one binding fast reaction model, and our models as $n$ binding slow reaction models. 

First, we study the stability of the one-binding slow reaction ODE system, and compare it with the fast reaction model. We obtain similar equilibrium states, but in terms of the parameter of the slow reaction model. Also, we derive the conditions, especially regarding the range of the CAR T-cell expansion rate parameter $p$, i) the minimum level of CAR T-cell expansion rate that provides the possibility that the CAR T-cell treatment can be successful, and ii) the level of CAR T-cell expansion rate that guarantees the escape from reaching the maximum tumor size.

Using our model, we then compare the fast and slow reaction model numerically, and show that the slow reaction model describes the experimental data more accurately, especially when the glioma antigen receptor density is low and/or mixed with relatively small numbers of CAR T-cells. In addition, we use the one-binding to five-binding slow reaction models to simulate low to high receptor density glioma cells, and study the assumption in the reaction rates that yields desired outcome, the decay and saturation of tumor size with respect to the antigen density levels. We come up with two different hypotheses to describe their connections, where the first hypothesis considers homogeneous rates across different numbers of CAR T-cell binding conjugates. More precisely, we assume identical rates for reactions involving glioma cells dying from the conjugates $I_j$. We show that the second hypothesis, where we consider non-homogeneous reaction rates, in particular, increasing tumor killing rates  as the number of bindings increases, reflects the reduced but saturated tumor size.

One of our future works is to calibrate the reaction rate parameters from experimental data, and verify whether our assumptions in hypothesis 2 on the reaction rates of multiple binding  model is valid. Currently, our model is calibrated to the time series data of glioma cells, and additional CAR T-cell dynamics data and experiments can be used to infer the reaction rates from data. We also hope to validate the effectiveness of the slow reaction model to in vivo data where we expect the interaction between the glioma cells and CAR T-cells to be slower than our current in vitro data. 
We propose to study and model the interaction of glioma cells and CAR T-cells with other immune cells as well. Modeling the interaction with patient's own immune cells including non CAR T- and B-cells, natural killer cells, macrophages, neutrophils, and dendritic cells, along with  cytokines may help understand how and why CAR T-cell conjugates are created as well as the role of other cells in the immune system in this process.

\printcredits


\bibliographystyle{cas-model2-names}

\bibliography{reference.bib}


\section*{Appendix}
\subsection*{A. Computation of two-binding model with fast reaction}
Our assumption in this case is that 
\begin{align}
    \dot{I_1}&=k_1^{(1)}CT-(k_{-1}^{(1)}+k_3^{(1)}+k_2^{(1)})I_1-k_1^{(2)}I_1T \nonumber \\ 
    &+k_{-1}^{(2)}I_2 =0\label{eq:2bindfast_1} \\ 
    \dot{I_2}&=k_1^{(2)}I_1T-(k_{-1}^{(2)}+k_4^{(2)}+k_3^{(2)}+k_2^{(2)})I_2=0
    \label{eq:2bindfast_2}
\end{align}
which means that the two conjugates decompose immediately right before the other CAR T-cells come to react with the conjugates themselves. By rearranging equation (\ref{eq:2bindfast_1}), we obtain
\begin{align}
    I_1&=\frac{k_1^{(1)}CT+k_{-1}^{(2)}I_2}{k_{-1}^{(1)}+k_3^{(1)}+k_2^{(1)}+k_1^{(2)}T} \nonumber \\
    &=\frac{k_1^{(1)}CT}{k_3^{(1)}+k_2^{(1)}+k_1^{(2)}T}\label{eq:conjugate1}
\end{align}
and by rearranging equation (\ref{eq:2bindfast_2}), we obtain
\begin{align}
    I_2&=\frac{k_1^{(2)}I_1T}{k_{-1}^{(2)}+k_4^{(2)}+k_3^{(2)}+k_2^{(2)}} \nonumber \\
    &=\frac{k_1^{(2)}I_1T}{k_4^{(2)}+k_3^{(2)}+k_2^{(2)}}\label{eq:conjugate2}
\end{align}
where we further assume that $k_{-1}^{(1)}\approx 0$. 
Now we look at $\dot{T}$
\begin{align*}
    \dot{T}&=pT\frac{C}{g+C}-\theta T-k_1^{(1)}CT+k_3^{(1)}I_1\\
    &-k_1^{(2)}I_1T+(2k_4^{(2)}+k_3^{(2)})I_2\\
    &=pT\frac{C}{g+C}-\theta T-k_1^{(1)}CT+k_3^{(1)}I_1\\
    &+\frac{k_4^{(2)}k_1^{(2)}-k_1^{(2)}k_3^{(2)}}{k_2^{(2)}+k_3^{(2)}+k_4^{(2)}}TI_1
\end{align*}
by substituting (\ref{eq:conjugate1}) into our equation and simplify it, we obtain
\begin{align}
    \dot{T}&=pT\frac{C}{g+C}-\theta T-\frac{k_1^{(1)}k_2^{(1)}}{k_1^{(2)}T+k_2^{(1)}+k_3^{(1)}}CT \nonumber \\
&-\frac{k_1^{(1)}k_1^{(2)}(k_2^{(2)}+2k_3^{(2)})}{(k_2^{(2)}+k_3^{(2)}+k_4^{(2)})(k_1^{(2)}T+k_2^{(1)}+k_3^{(1)})}CT^2
\label{eq:fastbinding_t}\end{align}
For $\dot{C}$, by using (\ref{eq:conjugate1}) and (\ref{eq:conjugate2}), we have
\begin{align*}
    \dot{C}&=\rho (1-\frac{C}{K})-k_1^{(1)}CT+k_2^{(1)}I_1+(k_3^{(2)}+k_2^{(2)})I_2\\
    &=\rho C(1-\frac{C}{K})-k_1^{(1)}CT+k_2^{(1)}I_1\\
    &+(k_3^{(2)}+k_2^{(2)})\frac{k_1^{(2)}I_1T}{k_4^{(2)}+k_3^{(2)}+k_2^{(3)}}\\
    &=\rho C(1-\frac{C}{K})-k_1^{(1)}CT\\
    &+(k_2^{(1)}+\frac{k_1^{(2)}(k_2^{(2)}+k_3^{(2)})}{k_4^{(2)}+k_3^{(2)}+k_2^{(2)}}T)I_1\\
    &=\rho C(1-\frac{C}{K})-k_1^{(1)}CT\\
    &+(k_2^{(1)}+\frac{k_1^{(2)}(k_2^{(2)}+k_3^{(2)})}{k_4^{(2)}+k_3^{(2)}+k_2^{(2)}}T)\frac{k_1^{(2)}CT}{k_3^{(1)}+k_2^{(1)}+k_1^{(2)}T}
\end{align*}
which can be rewritten as 
\begin{align}
    \dot{C}&=\rho C(1-\frac{C}{K})-\frac{k_1^{(1)}k_3^{(1)}}{k_2^{(1)}+k_3^{(1)}+k_1^{(2)}T}CT \nonumber \\ 
&-\frac{k_{4}^{(2)}k_1^{(2)}k_1^{(1)}}{(k_2^{(2)}+k_3^{(2)}+k_4^{(2)})(k_2^{(1)}+k_3^{(1)}+k_1^{(2)}T)}CT^2
\label{eq:fastbinding_c}
\end{align}
Therefore, (\ref{eq:fastbinding_t}) and (\ref{eq:fastbinding_c}) give our two-binding model with fast reaction.

\subsection*{B. Stability analysis of one-binding model with  slow reaction}
\label{sec:appenB}

The equilibrium points can be computed by:
\begin{align*}
p T \frac{C}{g + C}  - \theta  T - k C T + \alpha I_1&=0 \\ 
\rho  C ( 1-  \frac{ C}{K}) - k C T  + \beta I_1 &=0\\
 k C T -\gamma I_1&=0.   
\end{align*}
Observe that because of $kCT-\gamma I_1=0$, this ODE system should have similar equilibrium points as what we had in previous section. More precisely, using the relation $\frac{k}{\gamma}CT=I_1$, we can rewrite the three dimensional ODE system into a two dimensional ODE system: 
\begin{align*}
    & pT\frac{C}{g+C}-\theta T-kCT+\frac{\alpha k}{\gamma}CT=0 \\
    & \rho C(1-\frac{C}{K})-kCT+\frac{\beta k}{\gamma}CT=0
\end{align*}
The first equation gives us
$$T_{1,2}=\frac{\rho (1-\frac{C_{1,2}}{K})}{k-\frac{\beta k}{\gamma}},$$
and 
\begin{eqnarray*}
C_{1,2}&=&\frac{(p-\theta -kg+\frac{\alpha kg}{\gamma})}{2(k-\frac{\alpha k}{\gamma})} \\ 
&\,& \pm \frac{ \sqrt{(p-\theta-kg+\frac{\alpha kg}{\gamma})^2-4(\frac{\alpha k}{\gamma}-k)(-\theta g)}}{2(k-\frac{\alpha k}{\gamma})}.
\end{eqnarray*}
If we let $m:=k-\frac{\alpha k}{\gamma}$ and $n:=k-\frac{\beta k}{\gamma}$, these agree with $(T_1,C_1)$ and $(T_2,C_2)$ obtained with the fast reaction model. Therefore, again we have 4 equilibrium points 
$$(0,0,0), \quad (0,K,0),\quad (T_1,C_1,\frac{k}{\gamma} C_1T_1),\quad (T_2,C_2,\frac{k}{\gamma} C_2T_2)$$
We then notice the Jacobian matrix of the ODE system Eq. \eqref{eq:1bindslow} is 
\[
J(T,C,I_1) = \begin{pmatrix}
p\frac{C}{g+C}-\theta-kC & \frac{pgT}{(g+C)^2}-kT & \alpha \\
-kC & \rho-2\rho\frac{C}{K}-kT & \beta\\
kC & kT & {-\gamma}
\end{pmatrix}
\]
Let us examine each equilibrium points and compare them with the fast reaction model. 

1. The equilibrium point $(T,C,I_1) = (0,0,0)$ is the tumor-free and CAR T-cellfree case. The Jacobian matrix becomes 
$$
J(0,0,0) =  \begin{pmatrix}
    -\theta & 0 & \alpha \\
    0 & \rho & \beta\\
    0 & 0 & -\gamma
        \end{pmatrix}
$$
The corresponding characteristic polynomial will be 
$$\det \begin{pmatrix}
-\theta-\lambda & 0 & \alpha\\
0 & \rho-\lambda & \beta \\
0 & 0 & -\gamma-\lambda
\end{pmatrix}$$
Because this is an upper triangular matrix, the eigenvalues can be observed easily, i.e. $\lambda_1=-\theta, \lambda_2=\rho, \lambda_3=-\gamma$. We conclude that $(0,0,0)$ is an unstable saddle point.

2. The equilibrium point $(T,C,I_1) = (0,K,0)$ is the tumor reaching the maximal capacity with no CAR T-cell surviving. The Jacobian matrix becomes 
$$
J(0,K,0) = \begin{pmatrix}
\frac{pK}{g+K}-\theta-Kk & 0 &\alpha \\
-Kk & -a & \beta\\
Kk & 0 & -\gamma
\end{pmatrix}
$$
For simplicity, let $g(C):=p\frac{C}{g+C}-\theta-kC$, so $g(\frac{1}{b})=\frac{pK}{g+K}-\theta-Kk$.
The corresponding characteristic polynomial will be 
\begin{align*}
    P(\lambda)&=\det \begin{pmatrix}
    g(K)-\lambda & 0 &\alpha \\
-Kk & -a-\lambda & \beta\\
Kk & 0 & -\gamma-\lambda
    \end{pmatrix}\\
    &=-Kk\alpha (-\rho-\lambda)+(-\gamma -\lambda)(gK-\lambda)(-\rho-\lambda)
\end{align*}
By setting $P(\lambda)=0$, we obtain the first eigenvalue $\lambda_1=-\rho$, and the following quadratic equation, 
\begin{align*}
    -Kk\alpha +(-\gamma-\lambda)(g(K)-\lambda)=0. 
\end{align*}
This gives two additional solutions
\begin{equation*}
\lambda=\frac{g(K)-\gamma \pm \sqrt{(-g(K)+\gamma)^2+4(Kk\alpha+\gamma g(K))}}{2}.
\end{equation*}
Note that the expression inside the square root is always positive since 
\begin{equation*}
        (-g(K)+\gamma)^2+4(Kk\alpha+\gamma g(K))= (\gamma+g(K))^2+4Kk\alpha>0. 
\end{equation*}
Therefore, these two eigenvalues are always real. For this equilibrium point to be stable, we need the eigenvalues to be negative, i.e. 
\begin{eqnarray*}
g(K)-\gamma +\sqrt{(-g(K)+\gamma)^2+4(Kk\alpha+\gamma g(K))}<0,
\end{eqnarray*}
or equivalently, 
\begin{equation}
p<(\frac{g}{K}+1)(-\frac{k\alpha}{\gamma/K}+\theta+Kk)
\end{equation}
which comes from  
$\gamma g(K) <- Kk \alpha $. 
If Eq. \eqref{eq:p_cond2_slow} is satisfied, the equilibrium point $(0,K,0)$ is stable. Therefore, similar to \eqref{eq:p_cond2}, this condition provides the minimal level of CAR T-cell expansion rate to guarantee the escape from the scenario of reaching maximal cancer size.

3. The equilibrium points $(T_{1,2},C_{1,2},\frac{k}{\gamma}C_{1,2}T_{1,2})$ include $(T_1,C_1)$ that can be denoted as the CAR T-cell therapy success case by ordering the points as $0<C_1<C_2$ and $T_1>T_2>0$. For these equilibrium points to exist, we can obtain the two conditions, $ p-\theta-kg+\frac{\alpha kg}{\gamma} \geq  0$, and $(p-\theta-kg+\frac{\alpha kg}{\gamma})^2+4(\frac{\alpha k}{\gamma}-k)(\theta g) \geq 0,$ that reduces to 
\begin{equation}
\left( \sqrt{\theta} + \sqrt{g\left({\alpha k}/{\gamma}-k\right)}   \right)^2 \leq p . 
\end{equation}
This condition provides a minimum level of CAR T-cell expansion rate that makes treatment success possible. To examine the stability, 
let us define $ h(C) = - \theta + p\frac{C}{g+C} - kC + \frac{\alpha k}{\gamma}C $. 
The characteristic polynomial becomes 
\begin{eqnarray*}P(\lambda)
&= 
\textrm{det} \begin{pmatrix}
 -\frac{\alpha k}{\gamma} C_e -\lambda & h'(C_{e}) T_{e} - \frac{\alpha k}{\gamma} T_e & \alpha \\
-kC_{e} & \rho-2\frac{\rho}{K}C_{e}-kT_{e} -\lambda & \beta \\
kC_{e} & kT_{e} & -\gamma -\lambda
\end{pmatrix} \\ 
&=-\lambda^3+A\lambda^2+B\lambda+C
\end{eqnarray*}
where the coefficients are computed as following 
\begin{eqnarray*}
A=\rho-2\rho\frac{ C_e}{K}-\frac{\alpha k}{\gamma}C_e-kT_e-\gamma \hspace{3.63cm} \\
B=(\rho-2 {\rho}\frac{C_e}{K})\left( \gamma+ \frac{\alpha {k} C_e}{\gamma}\right) +(\beta-\gamma)kT_e - k h'(C_e)T_e C_e \\
C= {kC_e h'(C_e)T_e(\beta-\gamma)}. \hspace{4.92cm} 
\end{eqnarray*}
The stability of the equilibrium points will depend on the roots of this polynomial. Although it is difficult to analyze the condition from this polynomial, we know that for the CAR T-cell therapy success case $(T_1,C_1)$, $C<0$, since  $\beta = k_{-1}^{(1)}+k_2^{(1)}$ < $\gamma = k_{-1}^{(1)}+k_2^{(1)}+k_3^{(1)}$ and $h'(C_1)<0$, which provide us that $(T_1,C_1)$ is either stable or a saddle.

\end{document}